\documentclass[lineno]{biometrika}

\usepackage{amsmath, mathtools, amssymb}
\makeatletter
\renewcommand{\eqref}[1]{equation~\textup{\tagform@{\ref{#1}}}}
\makeatother
\usepackage{subcaption}

\usepackage{multirow}

\usepackage{newtxtext}
\usepackage[subscriptcorrection]{newtxmath}

\usepackage{natbib}

\usepackage{xcolor}
\usepackage{hyperref}
\hypersetup{
    colorlinks,
    linkcolor={red!50!black},
    citecolor={blue!50!black},
    urlcolor={blue!80!black}
}
\usepackage{url}
\usepackage[plain,noend]{algorithm2e}

\makeatletter
\renewcommand{\algocf@captiontext}[2]{#1\algocf@typo. \AlCapFnt{}#2} % text of caption
\def\@algocf@capt@plain{top}
\renewcommand{\algocf@makecaption}[2]{%
  \addtolength{\hsize}{\algomargin}%
  \sbox\@tempboxa{\algocf@captiontext{#1}{#2}}%
  \ifdim\wd\@tempboxa >\hsize%     % if caption is longer than a line
    \hskip .5\algomargin%
    \parbox[t]{\hsize}{\algocf@captiontext{#1}{#2}}% then caption is not centered
  \else%
    \global\@minipagefalse%
    \hbox to\hsize{\box\@tempboxa}% else caption is centered
  \fi%
  \addtolength{\hsize}{-\algomargin}%
}
\makeatother

\usepackage{xcolor}
\usepackage{booktabs}
\usepackage{longtable}
\usepackage{bbm}
\usepackage{ulem}
\renewcommand{\P}{\mathsf{P}}
\newcommand{\E}{\mathsf{E}}

\begin{document}
\nolinenumbers
\jname{Biometrika}
%% The year, volume, and number are determined on publication
\jyear{}
\jvol{}
\jnum{}
\cyear{}
%% The \doi{...} and \accessdate commands are used by the production team
%\doi{10.1093/biomet/asm023}
\accessdate{}

%% These dates are usually set by the production team
\received{d M yyyy}
\revised{d M yyyy}

%% The left and right page headers are defined here:
\markboth{M. Lin, G. Rempala, E. Kenah, and Q. Lin}{Exchangeable bootstrap and box calibration}

%% Here are the title, author names and addresses
\title{Simultaneous confidence bands for cumulative hazard via exchangeable bootstrap and box calibration} % all lower case

% , , Columbus, Ohio, U.S.A. 
\author{Min Lin}
\affil{Division of Biostatistics, College of Public Health, The Ohio State University,\\ Columbus, Ohio 43210, U.S.A.
\email{lin.5267@osu.edu}}

\author{Grzegorz Rempala}
\affil{Division of Biostatistics, College of Public Health, The Ohio State University,\\ Columbus, Ohio 43210, U.S.A.
\email{rempala.3@osu.edu}}

\author{Eben Kenah}
\affil{Division of Biostatistics, College of Public Health, The Ohio State University,\\ Columbus, Ohio 43210, U.S.A.
\email{kenah.1@osu.edu}}

\author{\and Qianying Lin}
\affil{Division of Biostatistics, College of Public Health, The Ohio State University,\\ Columbus, Ohio 43210, U.S.A. 
\email{lin.5268@osu.edu}}

\maketitle

\begin{abstract}
% \gr{Resampling-based simultaneous confidence bands for cumulative hazard functions often exhibit substantial finite-sample undercoverage under right censoring. We show that this problem has two distinct sources: the bootstrap resampling scheme may not accurately reproduce the finite-sample law of the Nelson–Aalen estimator, and standard grid-based calibration ignores between-jump excursions arising when a step estimator is compared with a continuous target function. To address these issues, we propose a bootstrap-based procedure that combines exchangeable resampling with a box-calibrated discrepancy. The exchangeable bootstrap preserves the ratio structure of the Nelson–Aalen estimator, while box calibration accounts for vertical deviations between successive event times. We establish conditional weak convergence of the exchangeable bootstrap, prove first-order equivalence between box and grid calibration, and show asymptotically nominal coverage of the resulting bands. Simulation studies across multiple hazard shapes and censoring levels demonstrate substantially improved finite-sample coverage, with the proposed method typically closest to the nominal level among the procedures considered. The method works directly on the cumulative-hazard scale and permits inference from time zero.}
Resampling-based simultaneous confidence bands for cumulative hazard functions often undercover in finite samples with right censoring.
We study two aspects of the construction that can contribute to this gap, the \emph{resampling scheme} and the \emph{calibration statistic}, and propose a procedure that intervenes on both.
The exchangeable bootstrap reweights the numerator and the denominator of the Nelson--Aalen ratio, preserving its ratio structure.
% \ml{The box-calibrated discrepancy constructs lower and upper step envelopes from adjacent values of the original and resampled Nelson--Aalen estimators and measures the largest vertical deviation from these envelopes.}
The box-calibrated discrepancy constructs lower and upper step envelopes from adjacent values of the original and resampled Nelson--Aalen estimators and measures the resulting vertical discrepancy.
We establish conditional weak convergence of the exchangeable bootstrap, prove that box calibration is first-order asymptotically equivalent to grid calibration, and show that the resulting band attains nominal coverage asymptotically. 
The box correction uses the same bootstrap paths and event-time grid as grid calibration; after each bootstrap path is formed, it requires only an additional linear pass over the event-time grid and therefore has negligible computational overhead.
% Diagnostics then trace the finite-sample undercoverage to the distributional channel: the calibrated statistic is a maximum over the event-time grid, and the bootstrap under-disperses this maximum, a finite-sample feature of approximating the sampling law of a maximum over a grid whose size grows with the sample.
% Box calibration acts as a geometric construction that compensates this distributional deficit by re-injecting the Nelson--Aalen jump scale.
In simulations across a range of hazard shapes and censoring levels, the exchangeable bootstrap with box calibration is, in most configurations, closest to nominal coverage among the methods considered.
A notable consequence is a ranking reversal: 
% the scheme that most faithfully reproduces the sampling law undercovers most under grid calibration, yet ranks best under box calibration;
% in our simulations this scheme is also closest to the nominal level.
the ratio-preserving exchangeable bootstrap has the lowest coverage under grid calibration, yet is usually closest to the nominal level after box calibration.
% \ml{Simulation diagnostics support an explanation based on the distribution of the grid maximum: in the settings considered, the grid-bootstrap critical value is below its grid-oracle counterpart, while box calibration adds a local jump-scale correction that often offsets this deficit for the exchangeable bootstrap.}
A melanoma data example illustrates the practical effect on the cumulative hazard bands.
The proposed procedure operates on the original cumulative-hazard scale, requires no variance-stabilizing transformation, and permits inference from time zero.
\end{abstract}


\section{Introduction}
\label{s:intro}
Simultaneous confidence bands are widely used to quantify uncertainty for cumulative hazard and survival curves under right censoring.
In the standard right-censoring setting, the estimand is a continuous monotone function, such as the cumulative hazard $H_0$ or the survival function $S_0$, while the corresponding estimator is a monotone step function, such as the Nelson--Aalen or Kaplan--Meier estimator \citep{nelson1969hazard,nelson1972theory,aalen1978nonparametric,kaplan1958nonparametric}.
Resampling-based simultaneous bands are often calibrated from the supremum distance between a resampled step estimator and the original step estimator, evaluated on the observed event-time grid \citep{akritas1986bootstrapping,lin1994confidence,bluhmki2019wild}.
However, as we demonstrate in this paper, currently used resampling-based simultaneous confidence band procedures for whole-interval inference on $[0,\tau]$ can exhibit severe finite-sample undercoverage. In particular, grid-based calibration may produce systematic undercoverage at practically relevant sample sizes, as demonstrated in Section~\ref{s:sim} below.
% For whole-interval inference on $[0,\tau]$, this grid-based calibration can produce systematic undercoverage at practically relevant sample sizes, as we demonstrate in Section~\ref{s:sim}.
% \sout{This undercoverage is not merely a matter of degree. Among grid-calibrated bands, the resampling scheme whose conditional law most faithfully reproduces the sampling law of the Nelson--Aalen estimator covers the worst, and the ranking of the three schemes reverses once the calibration is corrected.} 
The simulation results show more than a uniform change in coverage: the ranking among the three resampling schemes considered can reverse when the calibration statistic is changed.
This interaction is our main finding: how good a resampling scheme looks for simultaneous-band coverage depends on the calibration statistic used to judge it.

Early simultaneous bands for the Nelson--Aalen estimator used Brownian-bridge approximations to derive analytical critical values, with bands often formed on log- or arcsin-transformed cumulative-hazard scales \citep{hall1980confidence,nair1984confidence,bie1987confidence}.
A transformation can stabilize the variance near the boundary, but its finite-sample performance depends on the unknown data-generating mechanism; \citet{borgan1990note} compared several choices and found that no single transformation is uniformly preferred.
Transformations involving $\log$ are also undefined at the origin, which forces inference to begin at a positive time point.
Bootstrap-based calibration avoids choosing a transformation by estimating the quantile of the supremum statistic from resampled data.
\citet{akritas1986bootstrapping} established consistency of the classical bootstrap for the Kaplan--Meier estimator.
\citet{lin1994confidence} proposed a computationally convenient multiplier approach in which standard normal variates perturb the martingale increments of the Nelson--Aalen estimator.
\citet{sachs2022confidence} provides a recent overview of band construction methods in survival analysis, and \citet{dietrich2025wild} develops a general wild bootstrap framework based on martingale central limit theorems.
For pointwise inference at a fixed time, \citet{strawderman_accurate_1997} obtained second-order-correct confidence limits for the cumulative hazard using Efron's bootstrap with bias correction, but a comparable higher-order theory for simultaneous bands is not available.

We identify two aspects of the grid-calibrated bootstrap that the proposed procedure intervenes on.
The first concerns the \emph{resampling scheme}: the conditional law of the bootstrap may not closely approximate the sampling law of the Nelson--Aalen estimator at a given finite sample size $n$.
The multiplier approach of \citet{lin1994confidence} perturbs only the numerator of the Nelson--Aalen ratio, effectively bootstrapping the asymptotic linear term rather than the estimator itself.
The exchangeable bootstrap \citep{praestgaard1993exchangeably}, of which Efron's bootstrap \citep{efron_bootstrap_1979,efron_censored_1981} is a special case, reweights both the numerator and the denominator of the Nelson--Aalen ratio.
Thus the bootstrap estimator keeps the ratio structure of the original estimator.
% For renormlized standard exponential weights, this construction also yield a simple beta jump-sampling representation, which we use in the simulations.

The second aspect concerns the \emph{calibration statistic}.
Even with an excellent bootstrap approximation, a grid-calibrated statistic compares two step functions that jump at the same observed times, so the supremum reduces to a maximum over those times.
This misses the vertical gap between a step estimator and the continuous target just before or after a jump, a mismatch shared by every grid-calibrated resampling scheme.

The correction has a classical precedent. The Kolmogorov--Smirnov statistic checks both sides of each empirical-distribution jump, capturing the worst deviation within each inter-jump interval \citep{kolmogorov1933sulla,dvoretzky1956asymptotic}, and \citet{bickel1989confidence} showed a bootstrap analog to be asymptotically valid. For Nelson--Aalen-type bands, by contrast, existing bootstrap procedures take the supremum of the resampled minus the original estimator on the shared event-time grid, so it is attained on that grid with no between-jump correction \citep{andersen1993statistical,lin1994confidence,bluhmki2019wild}.

We intervene on both aspects simultaneously.
We use the exchangeable bootstrap for the Nelson--Aalen estimator and establish its conditional weak convergence, which has not, to our knowledge, been stated for the Nelson--Aalen ratio.
We then replace the grid-calibrated statistic with a \emph{box-calibrated} bootstrap statistic that constructs lower and upper step envelopes from adjacent values of the original and resampled Nelson--Aalen estimators and measures the resulting vertical discrepancy.
We prove that box calibration is first-order asymptotically equivalent to grid calibration, so the correction is not visible in the first-order limit, and that the combined procedure achieves asymptotically nominal coverage. 
Box calibration leaves the bootstrap law on the observed grid unchanged, but adds a local jump-scale correction using values already computed; after each bootstrap path has been formed, this requires only an additional linear pass over the same event-time grid.
% \sout{A simulation-based decomposition locates the undercoverage in the distributional channel: the grid statistic is a maximum over the event-time grid, and the bootstrap reproduces the upper tail of this maximum poorly to match its sampling law \citep{chernozhukov2017central,koike2026high}.
Simulations comparing exchangeable, weird, and wild bootstrap bands, each with grid and box calibration, together with the classical Hall--Wellner band \citep{hall1980confidence,andersen1993statistical}, show that box calibration reduces undercoverage and that the exchangeable bootstrap combined with box calibration is, in most configurations, closest to the nominal level. 
A melanoma data example illustrates how the proposed band behaves on the original cumulative-hazard scale and permits inference from time zero.

The remainder of this paper is organized as follows.
Section~\ref{s:model} sets up the random censorship model and recalls the Gaussian limit of the Nelson--Aalen process.
Section~\ref{s:xboot} introduces the exchangeable bootstrap, establishes its conditional weak convergence, and records the beta jump-sampling representation for exponential weights.
Section~\ref{s:method} defines the grid-calibrated benchmark and the box-calibrated discrepancy, establishes first-order equivalence with grid calibration, and proves asymptotic nominal coverage.
Section~\ref{s:sim} reports the simulation study, Section~\ref{s:data-example} illustrates the method on the melanoma data of \citet{andersen1993statistical}, and Section~\ref{s:discuss} concludes with extensions and open questions.
All proofs are collected in the Supplementary Material.

\section{Random censorship model}
\label{s:model}
Let $T$ be a positive event time with an absolutely continuous cumulative hazard function $H_0$, and let $C$ be a positive censoring time that is independent of $T$. Associated with $(T,C)$, define the observed time and event indicator by
\begin{equation*}
    Z\coloneqq T\land C,\quad \delta\coloneqq \mathbbm 1\{T\le C\},
\end{equation*}
where $x\land y\coloneqq \min\{x,y\}$ and $\mathbbm 1\{\cdot\}$ denotes the indicator function. For $t\in\mathcal T\coloneqq [0,\tau]$, define
\begin{equation*}
    N(t)\coloneqq \delta\mathbbm 1\{Z\le t\},\quad     Y(t)\coloneqq \mathbbm 1\{Z\ge t\},\label{eq:NandY}
\end{equation*}
where $\tau>0$ is fixed and chosen so that $\E Y(\tau)=\P (Z\ge \tau)>0$. Thus $N$ is the individual counting process and $Y$ is the individual at-risk process. Under independent right censoring, the process
\begin{equation*}
    M(t)\coloneqq N(t)-\int_0^t Y(u)\,\text{d}H_0(u),\quad t\in\mathcal T,
\end{equation*}
is a mean-zero square-integrable martingale with respect to the natural filtration. Taking expectations yields
\begin{equation*}
    \E\{N(t)\}=\int_0^t \E Y(u)\,\mathrm{d}H_0(u),\quad t\in\mathcal T.
\end{equation*}
Because $u\mapsto \E Y(u)$ is nonincreasing and $\E Y(\tau)>0$, we have $\inf_{u\in\mathcal T}\E Y(u)=\E Y(\tau)>0$. Therefore $H_0$ is uniquely identified by
\begin{equation}
    H_0(t)=\int_0^t\frac{\mathrm d\E N(u)}{\E Y(u)},\quad t\in\mathcal T.\label{eq:H0}
\end{equation}

Let $\mathcal D_n\coloneqq \{(Z_i,\delta_i)\colon i=1,\ldots,n\}$ be an i.i.d.\ sample of $(Z,\delta)$, and for $t\in\mathcal T$ define
\begin{equation*}
    N_i(t)\coloneqq\delta_i\mathbbm 1\{Z_i\le t\},\quad Y_i(t)\coloneqq \mathbbm 1\{Z_i\ge t\},
\end{equation*}
together with the empirical averages
\begin{equation*}
    \bar N_n(t)\coloneqq \frac1n\sum_{i=1}^n N_i(t),\quad \bar Y_n(t)\coloneqq \frac1n\sum_{i=1}^nY_i(t).
\end{equation*}
The Nelson--Aalen estimator 
% \citep{nelson1969hazard,nelson1972theory,aalen1978nonparametric} 
is
\begin{equation}
    \widehat H_n(t)\coloneqq \int_0^t \frac{\mathrm d\bar N_n(u)}{\bar Y_n(u)},\qquad t\in\mathcal T,\label{eq:NA_estimator}
\end{equation}
which is the sample analogue of \eqref{eq:H0}. Moreover, since $\bar Y_n(\tau)\to\E Y(\tau)>0$ almost surely, \eqref{eq:NA_estimator} is well-defined on $\mathcal T$ almost surely for all sufficiently large $n$.
% It follows from the functional delta-method (Theorem~3.10.4 and Example~3.10.20 of \citet{van_der_vaart_weak_2023}) that
By standard martingale arguments (see, for example, Section~IV.1.2 of \citet{andersen1993statistical}), one has
\begin{equation}
    \sqrt{n}(\widehat H_n-H_0)\rightsquigarrow \mathbb Z,\label{eq:original_limit}
\end{equation}
where $\mathbb Z$ is a zero-mean Gaussian process on $\mathcal T$ with continuous
sample paths and covariance function
\begin{equation*}
    \E \mathbb Z(s)\mathbb Z(t)=\int_0^{s\land t}\frac{\mathrm{d}H_0(u)}{\E Y(u)},\quad s,t\in\mathcal T.
\end{equation*}
Here $\rightsquigarrow$ denotes weak convergence. 

The weak limit \eqref{eq:original_limit} is the starting point for classical Gaussian-process-based simultaneous confidence bands, including the Hall--Wellner and equal-precision type construction \citep{hall1980confidence,nair1984confidence}; see, for example, Sections~IV.1.3.2 and IV.1.3.3 of \citet{andersen1993statistical}.

\section{Exchangeable bootstrap}
\label{s:xboot}
This section addresses the first of the two aspects identified in Section~\ref{s:intro}: the choice of resampling scheme, which controls how closely the conditional law of the bootstrap process matches the sampling law of $\sqrt{n}(\widehat{H}_n - H_0)$.

For the bootstrap analysis, and especially for simultaneous confidence bands, it is natural to work under the uniform norm. Accordingly, in this section we view $N$, $Y$, and the Nelson--Aalen estimator introduced above as stochastic processes on $\mathcal T$ with sample paths in $\ell^\infty(\mathcal T)$, the Banach space of bounded real-valued functions on $\mathcal T$ equipped with the uniform norm
\begin{equation*}
    \|f\|_{\mathcal T}\coloneqq \sup_{t\in\mathcal T}|f(t)|.
\end{equation*}
Since $\ell^\infty(\mathcal T)$ is nonseparable, all weak convergence statements and probability statements below are understood in the Hoffmann--J{\o}rgensen sense \citep{van_der_vaart_weak_2023}; in particular, outer expectation and outer probability are used whenever measurability is not immediate. 
A brief review is given in the Supplementary Material.

We first propose an exchangeable bootstrap procedure  \citep{praestgaard1993exchangeably} that yields a consistent bootstrap approximation to the sampling law of $\sqrt n(\widehat H_n-H_0)$. For each $n$, let $(W_{n1},\ldots,W_{nn})$ be an exchangeable, nonnegative random vector, independent of the data. Denote the weighted averaged processes by
\begin{align*}
        \bar N_n^W(t)\coloneqq \frac1n\sum_{i=1}^nW_{ni}N_i(t),
        \quad
        \bar Y_n^W(t)\coloneqq \frac1n\sum_{i=1}^nW_{ni}Y_i(t).
\end{align*}
When the weights are integer-valued, this corresponds to resampling the pairs $(N_i,Y_i)$ with multiplicities $W_{ni}$; more generally, the weights need not be integer-valued. Let 
\begin{align}
    \widehat H_n^W(t)&\coloneqq \int_0^t\frac{\mathrm d\bar N^W_n(u)}{\bar Y_n^W(u)}\label{eq:weighted_NA}
\end{align}
be the weighted estimator based on the weighted observations, in analogy with \eqref{eq:NA_estimator}. 

Denote
\begin{equation}
    0=E_{n0}<E_{n1}<\cdots<E_{n,m_n}<\tau<E_{n,m_n+1},\label{eq:event_time}
\end{equation}
where $E_{n1},\ldots,E_{n,m_n}$ are the uncensored event times in $\mathcal T$, and $E_{n,m_n+1}$ denotes the next uncensored event time after $\tau$, allowing $E_{n,m_n+1}=\infty$. Since both $\widehat H_n$ and $\widehat H_n^W$ are c\`adl\`ag step functions, they share the same jump grid $E_{n1},\ldots,E_{n,m_n}$. Then
\begin{equation*}
     \widehat H_n^W(t)=\sum_{j\colon E_{nj}\le t}\frac{W_{n,\pi(j)}}{\sum_{i=1}^nW_{ni}\mathbbm 1\{Z_i\ge E_{nj}\}},
\end{equation*}
is a sum of self-normalized weighted increments, where $\pi(j)$ is the index such that $Z_{\pi(j)}=E_{nj}$. 

The next theorem shows that, under suitable conditions on the weights, the bootstrap law of $\sqrt n(\widehat H_n^W-\widehat H_n)$, conditional on the observed data, consistently estimates the sampling law of $\sqrt n(\widehat H_n-H_0)$; the common Gaussian limit $\mathbb Z$ links the two.
\begin{condition}\label{cond:1}
For each $n$, let $(W_{n1},\ldots,W_{nn})$ be an exchangeable, nonnegative random vector, independent of the data, satisfying $\sum_{i=1}^nW_{ni}=n$. Assume:
\begin{enumerate}
    \item $\sup_n\int_0^\infty\P(|W_{n1}-1|>t)^{1/2}\,\mathrm{d}t<\infty$.
    \item $n^{-1/2}\E \max_{1\le i\le n}|W_{ni}-1|\to 0$.
    \item $n^{-1}\sum_{i=1}^n(W_{ni}-1)^2\to 1$ in probability.
\end{enumerate}
\end{condition}
\begin{remark}
     A sufficient condition for items 1 and 2 of Condition~\ref{cond:1} is that $\sup_n \E|W_{n1}|^{2+\varepsilon}<\infty$ for some $\varepsilon>0$.
\end{remark}
 \begin{theorem}\label{thm:boot}
     Assume $\inf_{t\in\mathcal T}\E Y(t)>0$. For weights satisfying Condition~\ref{cond:1}, 
     \begin{equation}
         \sqrt{n}(\widehat H_n^W-\widehat H_n)\overunderset{\P}{W}{\rightsquigarrow}\mathbb Z.\label{eq:bootstrap_CLT}
     \end{equation}
 \end{theorem}
Here $\overunderset{\P}{W}{\rightsquigarrow}$ means conditional weak convergence with respect to the bootstrap weights, in outer probability under the data-generating law. That is, with probability tending to one over the original data, the conditional law of $\sqrt n(\widehat H_n^W-\widehat H_n)$ is close to the law of $\mathbb Z$ in $\ell^\infty(\mathcal T)$. The exact meaning of conditional weak convergence is defined in the Supplementary Material.
The result follows by combining the exchangeable bootstrap central limit theorem for the empirical process \citep[Theorem~3.7.13]{van_der_vaart_weak_2023} with the Hadamard differentiability of the integration-ratio functional that produces the Nelson--Aalen estimator; the contribution here is the verification of Condition~\ref{cond:1} for the Nelson--Aalen ratio and the treatment of the right boundary at $\tau$, with a full proof in the Supplementary Material.
A per-individual reweighting of the counting and at-risk processes by time-constant multipliers does not in general reproduce the covariance of the Nelson--Aalen increments, and for a linearized statistic this mismatch has been documented \citep{dobler2014bootstrapping,bluhmki2019wild}.
Here the weights act on the full nonlinear ratio functional rather than on its linearization, and routing them through the conditional functional delta method \citep[Theorem~3.10.11]{van_der_vaart_weak_2023} transfers the correct limiting covariance of $\sqrt n(\widehat H_n-H_0)$, which is the content of Theorem~\ref{thm:boot}.

We next discuss exchangeable weights that satisfy Condition~\ref{cond:1}. Efron's original bootstrap \citep{efron_bootstrap_1979} amounts to drawing a multinomial vector $(W_{n1},\ldots,W_{nn})$ with size $n$ and probability $(1/n,\ldots,1/n)$. Another option is renormalized i.i.d.\ weights. Let $\xi_1,\xi_2,\ldots$ be a sequence of i.i.d.\ positive random variables with $\E |\xi_1|^{2+\varepsilon}<\infty$ for some $\varepsilon>0$. Then the renormalized weights $W_{ni}=\xi_i/\bar\xi_n$ satisfy Condition~\ref{cond:1} \citep{praestgaard1993exchangeably}, where $\bar \xi_n\coloneqq n^{-1}\sum_{i=1}^n\xi_i$. If we define
\begin{equation*}
    \bar N_n^\xi(t)\coloneqq \frac1n\sum_{i=1}^n\xi_iN_i(t),\quad     \bar Y_n^\xi(t)\coloneqq \frac1n\sum_{i=1}^n\xi_iY_i(t),
\end{equation*}
then the common factor $\bar\xi_n^{-1}$ cancels in \eqref{eq:weighted_NA}, so that
\begin{equation*}
    \widehat H_n^W(t)=\int_0^t\frac{\mathrm d\bar N_n^\xi(u)}{\bar Y_n^\xi(u)}.
\end{equation*}
One may choose, for example, standard exponential variables or Poisson variables with mean 1. 

For the renormalized i.i.d.\ standard exponential weights, we can obtain closed-form formulas. Let $\xi_1,\xi_2,\ldots$ be a sequence of i.i.d.\ standard exponential variables, independent of the data. Denote the jump sizes as
\begin{gather*}
    \Delta\widehat H_n^W(E_{nj})\coloneqq\widehat H_n^W(E_{nj})-\widehat H_n^W(E_{n,j-1})=\frac{W_{n,\pi(j)}}{\sum_{i=1}^nW_{ni}\mathbbm 1\{Z_i\ge E_{nj}\}},\\
    \Delta\widehat H_n(E_{nj})\coloneqq\widehat H_n(E_{nj})-\widehat H_n(E_{n,j-1})=\frac{1}{R_{nj}},
\end{gather*}
where $R_{nj}\coloneqq n\bar Y_n(E_{nj})$ is the number of subjects at risk just before the $j$th uncensored event. By the exchangeability and stick-breaking, the exchangeable bootstrap samples jump sizes from
\begin{equation}
    \Delta\widehat H_n^W(E_{nj})\mid \mathcal D_n\sim \mathrm{Beta}\bigl(1,R_{nj}-1\bigr).\label{eq:beta_sampling}
\end{equation}
Thus, 
\begin{equation*}
    \E[ \Delta\widehat H_n^W(E_{nj})\mid \mathcal D_n]=\frac{1}{R_{nj}}=\Delta\widehat H_n(E_{nj}),\quad \mathsf{var}[ \Delta\widehat H_n^W(E_{nj})\mid \mathcal D_n]=\frac{R_{nj}-1}{R_{nj}^2(R_{nj}+1)}.
\end{equation*}
A similar beta sampling scheme has been studied by \citet{lo_bayesian_1993} for the Kaplan-Meier estimator. 

In contrast, \citet{lin1994confidence} proposed a Gaussian perturbation,
which in the present setting corresponds to a wild bootstrap that samples
jump sizes independently from
$R_{nj}^{-1}\mathrm{Normal}(1,1)$. These resampled jump sizes have
conditional mean $\Delta\widehat H_n(E_{nj})$ and conditional variance
$R_{nj}^{-2}$, but they are unbounded and may also be negative.
\citet{dobler2019confidence} considered alternative wild-bootstrap
weights leading to jump sizes distributed as
$R_{nj}^{-1}\mathrm{Poisson}(1)$ and $\mathrm{Exponential}(R_{nj})$; in their
simulation study, the exponential choice performed better than the
Poisson and Gaussian ones. Still, the Poisson and exponential jump sizes
are unbounded from above. By contrast, the weird bootstrap samples
independently from
$R_{nj}^{-1}\mathrm{Binomial}(R_{nj},R_{nj}^{-1})$
\citep{andersen1993statistical}, which is supported on $[0,1]$, with
conditional mean $\Delta\widehat H_n(E_{nj})$ and conditional variance
$R_{nj}^{-3}(R_{nj}-1)<R_{nj}^{-2}$. The beta sampling scheme in
\eqref{eq:beta_sampling} may be viewed as a smoothed version of the
weird bootstrap. It preserves the same conditional mean while having smaller conditional variance than both the weird bootstrap and the wild-bootstrap schemes
based on Gaussian, Poisson, or exponential weights.

\section{Benchmark bands and box calibration}
\label{s:method}
This section addresses the second aspect identified in Section~\ref{s:intro}: the calibration statistic.
We first record in Section~\ref{ssec:grid} the grid-calibrated benchmark band and exhibit its geometric mismatch with the oracle supremum against the continuous target, and then in Section~\ref{ssec:box} introduce the box-calibrated discrepancy that corrects this mismatch without altering the event-time grid.

\subsection{A grid-calibrated benchmark and its geometric mismatch}
\label{ssec:grid}
Before deriving a consequence for the uniform norm, we record the relevant measurability facts. Fixing the data, the weighted process $\widehat H_n^W-\widehat H_n$ is a c\`adl\`ag step function with jumps only at the observed uncensored event times. Hence it takes values in a finite-dimensional subspace of $\ell^\infty(\mathcal T)$, and, on the event that the weighted risk-set denominators are positive, it is a Borel measurable function of the weights. On the other hand, $\widehat H_n-H_0$ need not be Borel measurable as an $\ell^\infty(\mathcal T)$-valued map. Nonetheless, since $\widehat H_n$ has c\`adl\`ag sample paths, it is ball measurable under the uniform norm \citep{pollard1984convergence}, and therefore $\|\widehat H_n-H_0\|_{\mathcal T}$ is an ordinary measurable real-valued random variable. 

Define the supremum discrepancies by
\begin{equation}
    \Delta_n\coloneqq \|\widehat H_n-H_0\|_{\mathcal T},\quad 
    \widehat\Delta_{n,\mathrm{grid}}^W \coloneqq \|\widehat H_n^W-\widehat H_n\|_{\mathcal T}.
\end{equation}
For $\alpha\in(0,1)$, define the $(1-\alpha)$-quantile of the conditional distribution of $\widehat\Delta_{n,\mathrm{grid}}^W$ by
    \begin{equation}
    \hat q^W_{n,\mathrm{grid}}(1-\alpha)\coloneqq \inf\left\{x\in\mathbb R\colon \P\bigl(\widehat\Delta_{n,\mathrm{grid}}^W\le x\mid \mathcal D_n\bigr)\ge 1-\alpha\right\}.
\end{equation}
\begin{proposition}\label{prop:1}
    Assume the conditions of Theorem~\ref{thm:boot}. 
Then \begin{equation}
       \sup_{x\in\mathbb R}\left|\P\bigl(\widehat\Delta_{n,\mathrm{grid}}^W\le x\mid \mathcal D_n\bigr)-\P\bigl(\Delta_n\le x\bigr)\right|\overset{\P}{\to} 0.\label{eq:cor1}
    \end{equation} 
    Moreover, for each fixed $\alpha\in(0,1)$,
    \begin{equation}
        \P\!\left( \Delta_n\le \hat q_{n,\mathrm{grid}}^{W}(1-\alpha) \right) \to 1-\alpha.
    \end{equation}
\end{proposition}
 
% The exchangeable bootstrap scheme is easy to implement in practice: $\widehat H^W_n$ uses the same observed event-time grid as $\widehat H_n$, while only the weights are resampled. Given the observed data, the conditional distribution of $\widehat \Delta_{n,\mathrm{grid}}\mid \mathcal D_n$ can be obtained by Monte Carlo methods: resampling weights.

% Let $W^{(1)},\ldots, W^{(B)}$ be $B$ independent weight vectors satisfying Condition~\ref{cond:1}. 
% empirical distribution of 
% \begin{equation*}
%     \|\widehat H_n^{W^{(1)}}-\widehat H_n\|_{\mathcal T},\ldots,\|\widehat H_n^{W^{(B)}}-\widehat H_n\|_{\mathcal T}
% \end{equation*}
% approximates the bootstrap distribution; by Proposition~\ref{prop:1}, this in turn consistently approximates the sampling distribution of $\|\widehat H_n-H_0\|_{\mathcal T}$. 
By Proposition~\ref{prop:1}, a natural fixed-width $(1-\alpha)\%$ confidence band is obtained by calibrating the bootstrap supremum discrepancy on the shared jump grid:
\begin{equation}
    \widehat H_n \pm \hat q_{n,\mathrm{grid}}^{W}(1-\alpha).
    \label{eq:grid_band}
\end{equation}
This grid-calibrated band serves only as a benchmark, to expose the finite-sample mismatch addressed below.
% \gr{The key point is that the oracle estimation error compares a step estimator with a continuous target function, whereas the bootstrap estimator $\widehat \Delta^W_{n,\mathrm{grid}}$ compares two step functions defined on the same event-time grid. Consequently, the bootstrap discrepancy may fail to capture the largest between-jump excursions that contribute to the true supremum error.} 
The grid-calibrated discrepancy satisfies
\begin{equation}
    \widehat\Delta_{n,\mathrm{grid}}^W=     \max_{i=1,\ldots,m_n} \bigl|\widehat H_n^W(E_{ni})-\widehat H_n(E_{ni}) \bigr|.
    \label{eq:grid_diff}
\end{equation}
Thus \eqref{eq:grid_diff} checks only the vertical discrepancies at the common jump points. In contrast, the actual supremum error of the estimator against the continuous target $H_0$ is
\begin{equation}
    \begin{aligned}
        \Delta_n &=  \max_{i=1,\ldots,m_n} \max\!\big\{\widehat H_n(E_{ni})-H_0(E_{ni}), H_0(E_{ni}) - \widehat{H}_n(E_{n,i-1})\big\}\\
        &\qquad \vee \{H_0(\tau)-\widehat H_n(E_{m_n})\},
    \end{aligned}\label{eq:Delta_n}
\end{equation}
where $x\lor y\coloneqq \max\{x,y\}$. Indeed, on each interval $[E_{n,i-1},E_{ni})$ the estimator is constant at $\widehat H_n(E_{n,i-1})$, while $H_0$ is continuous and increasing, so the largest positive excursion of $H_0-\widehat H_n$ on that interval is attained at the right endpoint. Consequently, \eqref{eq:grid_diff} and \eqref{eq:Delta_n} are geometrically different in finite samples: the former only checks the shared grid, whereas the latter also records the vertical excursions between consecutive jumps and at the boundary point $\tau$.

\subsection{Box calibration}
\label{ssec:box}
We correct the geometric mismatch without changing the observed event-time grid by augmenting each interval with a vertical calibration.
% The key idea is to retain the same event-time partition \eqref{eq:event_time}, but enrich each interval by a vertical calibration. 
This leads to a box-calibrated discrepancy. 

\begin{figure}[ht!]
    \centering
    \includegraphics[width=.8\linewidth]{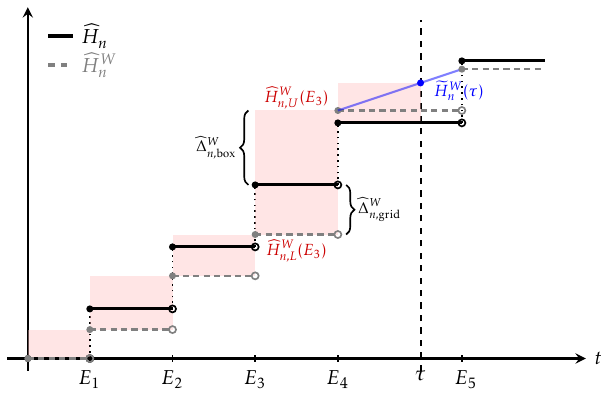}
    \caption{
        Illustration of the box calibration on a sample with uncensored event times $E_1,\ldots,E_5$ and $\tau\in(E_4, E_5)$.
        Horizontal solid black lines is the Nelson--Aalen estimator $\widehat{H}_n$; 
        horizontal dashed grey lines represent the bootstrap counterpart $\widehat{H}^W_n$ on the same event-time grid.
        As an illustrative example, $\widetilde{H}^W_n(\tau)$ is the interpolant obtained from a linear interpolation on the interval $[E_4,E_5)$ at the boundary $\tau$.
        As the shading illustrates, the step envelopes $\widehat{H}^W_{n,L}$ and $\widehat{H}^W_{n,U}$ bracket $\widehat{H}_n$ on each interval $[E_i, E_{i+1})$, together with the boundary value $\widetilde{H}^W_n(\tau)$ that fixes $\widehat{H}^W_{n,U}$ on the tail interval $[E_4,\tau]$;
        the box-calibrated discrepancy $\widehat{\Delta}^W_{n,\mathrm{box}}$ is the largest vertical deviation between $\widehat{H}_n$ and these envelopes, while $\widehat{\Delta}^W_{n,\mathrm{grid}}$ is the largest deviation between $\widehat{H}_n$ and $\widehat{H}_n^W$ across grids.
        In this illustration, both $\widehat{\Delta}^W_{n,\mathrm{box}}$ and $\widehat{\Delta}^W_{n,\mathrm{grid}}$ are achieved in $[E_3,E_4)$.
    }
    \label{fig:illustration}
\end{figure}

To make this explicit, define two c\`adl\`ag step functions $\widehat H_{n,L}^W$ and $\widehat H_{n,U}^W$ by
\begin{align*}
    \widehat H_{n,L}^W(E_{ni}) &\coloneqq \widehat H_n(E_{ni})\land \widehat H_n^W(E_{ni}), \quad i=0,\ldots,m_n,\\
    \widehat H_{n,U}^W(E_{ni}) &\coloneqq \widehat H_n(E_{ni})\lor \widehat H_n^W(E_{n,i+1}), \quad i=0,\ldots,m_n-1,
\end{align*}
and, for a value $K^W_{n,\tau}\in[\widehat H_n^W(E_{n,m_n}), \widehat H_n^W(E_{n,m_n+1})]$, define
\begin{equation*}
    \widehat H^W_{n,U}(E_{n,m_n})\coloneqq \widehat H_n(E_{n,m_n})\lor K^W_{n,\tau}.
\end{equation*}
Both $\widehat H_{n,L}^W$ and $\widehat H_{n,U}^W$ are c\`adl\`ag step functions, constant on each interval $[E_{ni},E_{n,i+1})$. The boundary value $K_{n,\tau}^W$ may be chosen by any measurable rule taking values in $[\widehat H_n^W(E_{n,m_n}),\,\widehat H_n^W(E_{n,m_n+1})]$. The box-calibrated discrepancy is then defined by
\begin{equation}
\begin{aligned}
     \widehat \Delta^W_{n,\mathrm{box}}&\coloneqq \|\widehat H_n-\widehat H_{n,L}^W\|_{\mathcal T}\vee \|\widehat H_{n,U}^W-\widehat H_n\|_{\mathcal T}\\
     &\ =\max_{i=1,\ldots,m_n} \max\!\big\{\widehat H_n(E_{ni})-\widehat H_n^W(E_{ni}), \widehat H_n^W(E_{ni}) - \widehat{H}_n(E_{n,i-1})\big\}\\
        &\qquad \vee \{K_{n,\tau}^W-\widehat H_n(E_{n,m_n})\}.
\end{aligned}\label{eq:box_discrepancy}
\end{equation}
Compared with \eqref{eq:grid_diff}, the event-time grid is unchanged. What is added is a vertical calibration within each interval. As a result, the discrepancy now matches the structure of \eqref{eq:Delta_n}. In this sense, the method still works on the same horizontal partition, but each
interval is now treated as a two-dimensional box rather than as a single point on the grid. By construction, 
\begin{equation*}
    \widehat H_{n,L}^W\le\widehat H_n\land \widehat H_n^W\le \widehat H_n\lor \widehat H_n^W\le \widehat H_{n,U}^W.
\end{equation*}
This explains the name \emph{box calibration}: on each interval $[E_{ni},E_{n,i+1})$, the two step envelopes $\widehat H_{n,L}^W$ and $\widehat H_{n,U}^W$ define the lower and upper edges of a vertical box around $\widehat H_n$. The calibrated discrepancy records the largest vertical deviation from these boxes. An illustration is shown in Fig.~\ref{fig:illustration}.

A convenient way to realize the boundary value $K_{n,\tau}^W$ is through a monotone continuous interpolation of $\widehat H_n^W$ between successive event times. Let $\widetilde H_n^W$ be any such interpolation. Then $\widetilde H_n^W(\tau)\in[\widehat H_n^W(E_{n,m_n}), \widehat H_n^W(E_{n,m_n+1})]$, and \eqref{eq:box_discrepancy} can be written as
% \begin{equation}
%     \widetilde H_n^W(t)\coloneqq \widehat H_n^W(E_{ni})+(t-E_{ni})\frac{\widehat H_n^W(E_{n,i+1})-\widehat H_n^W(E_{ni})}{E_{n,i+1}-E_{ni}}, \quad t\in[E_{ni},E_{n,i+1})\label{eq:linear_inter},
% \end{equation}
% for $i=0,\ldots,m_n$ whenever $E_{n,i+1}<\infty$. If $E_{n,m_n+1}=\infty$, set $\widetilde H_n^W(t)\coloneqq \widehat H_n^W(E_{n,m_n})$ for $t\in[E_{n,m_n},\tau]$. Then
\begin{equation*}
\begin{aligned}
    \widehat \Delta_{n,\mathrm{box}}^W&=\|\widetilde H_n^W-\widehat H_n\|_{\mathcal T}\\
    &=\max_{i=1,\ldots,m_n} \max\!\big\{\widehat H_n(E_{ni})-\widehat H_n^W(E_{ni}), \widehat H_n^W(E_{ni}) - \widehat{H}_n(E_{n,i-1})\big\}\\
        &\qquad \vee \{\widetilde H_n^W(\tau)-\widehat H_n(E_{n,m_n})\}.
\end{aligned}
\end{equation*}
Thus the resulting discrepancy depends on the interpolation only through the boundary value $K_{n,\tau}^W$ at $\tau$. 

For $\alpha\in(0,1)$, define the $(1-\alpha)$-quantile of the conditional distribution of the box-calibrated discrepancy by
    \begin{equation}
        \hat q^W_{n,\mathrm{box}}(1-\alpha)\coloneqq \inf\left\{x\in\mathbb R\colon \P\bigl(\widehat\Delta_{n,\mathrm{box}}^W\le x\mid \mathcal D_n\bigr)\ge 1-\alpha\right\}.
    \end{equation}
Since $\widehat\Delta_{n,\mathrm{box}}^W\ge \widehat\Delta_{n,\mathrm{grid}}^W$, we have $\hat q_{n,\mathrm{box}}^W(1-\alpha)\ge \hat q_{n,\mathrm{grid}}^{W}(1-\alpha)$ for each $\alpha\in(0,1)$. Hence the box-calibrated band 
\begin{equation}
    \widehat H_n \pm \hat q_{n,\mathrm{box}}^W(1-\alpha)\label{eq:box_band}
\end{equation}
is no narrower than the benchmark grid-calibrated band \eqref{eq:grid_band}. The next theorem shows that box calibration still approximates the oracle exact distribution of $\Delta_n$ while enlarging the bootstrap discrepancy to match the structure of the oracle error \eqref{eq:Delta_n}.

 \begin{theorem}\label{thm:box}
     Assume $\inf_{t\in\mathcal T}\E Y(t)>0$. For weights satisfying Condition~\ref{cond:1}, the box-calibrated discrepancy \eqref{eq:box_discrepancy} satisfies
         \begin{equation}
       \sup_{x\in\mathbb R}\left|\P\bigl(\widehat\Delta_{n,\mathrm{box}}^W\le x\mid \mathcal D_n\bigr)-\P\bigl(\Delta_n\le x\bigr)\right|\overset{\P}{\to} 0,
    \end{equation}
    and for each $\alpha\in(0,1)$,
    \begin{equation}
    \P\!\left( \Delta_n \le \hat q_{n,\mathrm{box}}^W(1-\alpha) \right) \to 1-\alpha.
\end{equation}
 \end{theorem}
 \begin{remark}
     Under the same assumptions, one can show that the box calibration also preserves the limit when applying to the weird and wild bootstraps.
 \end{remark}

% Box calibration is thus a geometric \emph{construction}: it enlarges each interval's discrepancy to the structure of the oracle error \eqref{eq:Delta_n}. By Theorem~\ref{thm:box} the enlargement is invisible in the first-order limit, so its effect is confined to finite samples. 
% We anticipate one point that the simulations of Section~\ref{s:sim} and the Supplementary Material make precise: the between-jump excursion that grid calibration omits contributes almost nothing to the quantile that sets coverage, so the finite-sample benefit of the enlargement is not the restoration of a geometric gap but the compensation of a distributional deficit in the bootstrap maximum, which we examine in Section~\ref{ssec:coverage_tab}.

The preceding construction leads to a simple Monte Carlo procedure, summarized in Algorithm~\ref{alg:xboot_box}.

\begin{algorithm}[htbp]
\caption{Exchangeable bootstrap with box calibration}
\label{alg:xboot_box}
\DontPrintSemicolon
\KwIn{Observed data $\mathcal D_n$, upper time point $\tau>0$, number of bootstrap replications $B$, and nominal error probability $\alpha\in(0,1)$.}
\KwOut{A simultaneous asymptotic $(1-\alpha)\%$ confidence band for $H_0$ on $[0,\tau]$.}

Compute the uncensored event times
$0=E_{n0}<E_{n1}<\cdots<E_{n,m_n}<\tau<E_{n,m_n+1}$
and the Nelson--Aalen estimator $\widehat H_n$ in \eqref{eq:NA_estimator}.\;

\For{$b=1,\ldots,B$}{
    Generate exchangeable bootstrap weights
    $W_{n1}^{(b)},\ldots,W_{nn}^{(b)}$ satisfying Condition~\ref{cond:1}.\;

    Construct the weighted Nelson--Aalen estimator
    $\widehat H_n^{W^{(b)}}$ in \eqref{eq:weighted_NA}
    on the same event-time grid.\;

    Choose a boundary value
    \[
        K_{n,\tau}^{W^{(b)}}\in
        \left[
        \widehat H_n^{W^{(b)}}(E_{n,m_n}),
        \widehat H_n^{W^{(b)}}(E_{n,m_n+1})
        \right],
    \]
    for example by linear interpolation at $\tau$.\;

    Compute the box-calibrated discrepancy
    $\widehat{\Delta}^{W^{(b)}}_{n,\mathrm{box}}$
    as in \eqref{eq:box_discrepancy}.\;
}

Let $\hat q_{n,B,\mathrm{box}}^W(1-\alpha)$ be the empirical
$(1-\alpha)$-quantile of
$\widehat{\Delta}^{W^{(1)}}_{n,\mathrm{box}},\ldots,
\widehat{\Delta}^{W^{(B)}}_{n,\mathrm{box}}$.\;

Return the band
\[
    \widehat H_n(t)
    \pm
    \hat q_{n,B,\mathrm{box}}^W(1-\alpha),
    \qquad 0\le t\le \tau.
\]
\end{algorithm}

\section{Simulation}
\label{s:sim}
\subsection{Settings}

We evaluated the finite-sample coverage of simultaneous confidence bands for
the cumulative hazard function under four event-time models: the standard
exponential distribution, a Weibull distribution with decreasing hazard, a
Weibull distribution with increasing hazard, and a piecewise-constant hazard.
For the Weibull$(k,b)$ models, we use the parameterization
\begin{equation*}
H_0(t)=(t/b)^k,
\end{equation*}
so that the standard exponential model corresponds to $(k,b)=(1,1)$, the
decreasing-hazard model to $(k,b)=(0.5,0.5)$, and the increasing-hazard model
to $(k,b)=(2,1/\Gamma(3/2))$. The piecewise-constant hazard is
\begin{equation}
h_0(t)=0.5\,\mathbbm{1}\{t\le \tau_0\}+2\,\mathbbm{1}\{t>\tau_0\},
\qquad \tau_0=2\log(3/2). \label{eq:piecewise_constant}
\end{equation}
These four event-time models were chosen so that $\E(T)=1$ in every case. For
each model, we considered light ($\approx 20\%$) and heavy ($\approx 50\%$)
right censoring, sample sizes $n\in\{15,25,50,100,200,500\}$, and nominal
confidence levels $100(1-\alpha)\%\in\{99\%,95\%,90\%\}$.

We compared seven methods: the exchangeable, weird, and wild bootstrap bands, each combined with grid or box calibration, together with the classical Hall--Wellner band. To simplify the tables and repeated comparisons below, we use the labels XBoot--Grid, XBoot--Box, Weird--Grid, Weird--Box, Wild--Grid, Wild--Box, and HW throughout this section.

We did not include Nair's equal-precision (EP) band as a main benchmark \citep{nair1984confidence}.
Although it is a classical survival-function band, it belongs to a different comparison class from fixed-width cumulative-hazard bands studied here.
The EP construction is a studentized survival-scale band, and its implementation depends on the chosen transformation and a positive lower-bound time.
These choices are useful in applications, but they introduce additional method-specific degrees of freedom that are not part of HW or of the proposed original-scale box-calibrated band.
We therefore use HW as the classical original-scale benchmark in the main simulation study, and use logHW in the data example (Section~\ref{s:data-example}) only to illustrate the effect of a standard transformed-scale construction.

For the exchangeable bootstrap, we used renormalized i.i.d.\ standard
exponential weights, which produce the beta sampling scheme in
\eqref{eq:beta_sampling}. For the wild bootstrap, we used i.i.d.\ standard
exponential weights. Box calibration was carried out by linear interpolation
between successive jump points. Empirical coverage was estimated from
$100{,}000$ Monte Carlo replications, using $2{,}000$ bootstrap resamples for
each of $\alpha=0.01,0.05,0.10$, with upper time bound fixed at $\tau=1$. The
corresponding 95\% Monte Carlo reference intervals, reported in percent, are
$[98.938,99.062]$, $[94.865,95.135]$, and $[89.814,90.186]$ for the 99\%,
95\%, and 90\% nominal levels, respectively.

For each event-time model, let $T$ denote the failure time and let $C$ denote
an independent censoring time. Under heavy censoring ($\approx 50\%$), $C$
was generated from the same distribution as $T$, so that $\P(C<T)=1/2$ by
symmetry. Under light censoring ($\approx 20\%$), $C$ was generated from the
same parametric family as $T$, with parameters chosen so that $\P(C<T)=1/5$.
Specifically, when $T$ follows a Weibull distribution with cumulative hazard
$H_0(t)=(t/b)^k$, the censoring time $C$ follows a Weibull distribution with
cumulative hazard
\begin{equation*}
    H_C(t)=\{t/(4^{1/k}b)\}^k,
\end{equation*}
which gives $\P(C<T)=1/5$. For the piecewise-constant hazard model, the
censoring distribution was generated from the proportional hazards model
\begin{equation*}
h_C(t)=\tfrac14 h_0(t),
\end{equation*}
which again gives $\P(C<T)=1/5$. The observed data for each subject consist of
$(T\land C,\mathbbm{1}\{T\le C\})$.

\subsection{Empirical coverage}
\label{ssec:coverage_tab}
All bootstrap schemes and calibration methods were implemented in the open-source \texttt{R} package \texttt{xbootboxR}, available at \url{https://github.com/MinLinSTAT/xbootboxR}, which was used to produce all simulation results reported below.

Table~\ref{tab:alpha005} reports the empirical coverage probabilities for the nominal 95\% level.
The corresponding results for the nominal 99\% and 90\% levels are reported in
Supplementary Tables~S1 and~S2, respectively.
Three findings emerge consistently across the hazard shapes and censoring levels.

First, box calibration increases coverage for every bootstrap scheme in every setting considered.
This direction is guaranteed by construction:
since $\widehat\Delta^W_{n,\mathrm{box}}\ge\widehat\Delta^W_{n,\mathrm{grid}}$ pointwise, the calibrated quantile satisfies $\widehat q^W_{n,\mathrm{box}}(1-\alpha) \ge \widehat q^W_{n,\mathrm{grid}}(1-\alpha)$, so the box band is never narrower and its coverage is never lower.
The empirical question is therefore the magnitude of the increase, and whether it corrects undercoverage or overshoots the nominal level.
At small sample sizes and heavy censoring, the increase is large and corrective:
at $n=25$ with $50\%$ censoring and $\alpha=0.05$, XBoot--Grid covers at $67.80$--$85.11\%$ depending on the hazard shape, whereas XBoot--Box covers at $92.47$--$95.67\%$.
For the already wider wild bootstrap bands, by contrast, the same box enlargement often produces overcoverage.
The difference between grid and box calibration diminishes as $n$ increases, and by $n=500$ the corresponding coverages are within $1$--$2$ percentage points of each other.

Second, the three bootstrap schemes exhibit a ranking reversal between grid and box calibration.
Among grid-calibrated methods, XBoot--Grid produces the lowest empirical coverage, followed by Weird--Grid, while Wild--Grid is usually closest to the nominal level.
This ordering agrees with the conditional-variance comparison in Section~\ref{s:xboot}: the exchangeable bootstrap has the smallest conditional variance for each jump increment, followed by the weird bootstrap, while the wild bootstrap has the largest.
Under box calibration, however, the ranking reverses in most configurations. 
XBoot--Box is usually closest to the nominal level, Wild--Box tends to exceed it, and Weird--Box falls in between the two.
At $\alpha=0.05$ with $n=25$ and heavy censoring under the $\mathrm{Exponential}(1)$ hazard, XBoot--Box covers at $95.29\%$, Weird--Box at $95.75\%$, and Wild--Box at $97.46\%$, against a $95\%$ Monte Carlo reference interval of $[94.865, 95.135]$.

The supplementary diagnostics clarify this reversal at the level of critical values. 
Section~S4 decomposes the expected $\sqrt{n}$-scaled box critical-value bias into a grid-distribution term $D$ and a net box-inflation term $G$. 
The diagnostics shows that $D$ is negative for all three bootstrap schemes, so the grid-bootstrap critical value is below its grid-oracle counterpart on average. 
The deficit is largest for the exchangeable bootstrap. The box-inflation term $G$ is positive for all three schemes and is largest for the wild bootstrap. 
For XBoot--Box, the negative grid-distribution term and the positive box-inflation term are often of comparable size; averaged over the four heavy-censoring hazards in the diagnostic experiment, the corresponding values are approximately $\bar D=-0.82$, $\bar G=0.76$, and $\bar b = \bar D + \bar G=-0.06$.
For Wild--Box, the positive box-inflation term is larger than the grid-distribution deficit, with approximately $\bar D=-0.40$, $\bar G=0.95$, and $\bar b=0.55$. Weird--Box lies between these two patterns.
Section~S5 further shows that, at fixed $n$, the magnitude of the negative grid-distribution term decreases as the event-time grid is thinned.
Together, these diagnostics support the interpretation that the ranking reversal reflects an interaction between the finite-sample distribution of the grid maximum and the jump-scale enlargement introduced by box calibration.

Third, the Hall--Wellner band exhibits undercoverage comparable to the grid-calibrated bootstrap methods at small $n$ and converges to nominal coverage at a similar rate.
At $n=500$, the Hall--Wellner coverage approaches but does not uniformly attain the nominal level, with residual undercoverage most pronounced at $\alpha=0.01$ and under heavy censoring with the decreasing-hazard $\mathrm{Weibull}(0.5,0.5)$ shape.

These patterns are remarkably stable across all four hazard functions, suggesting the finite-sample behavior is driven less by a particular hazard shape than by the interaction between the resampling scheme, the event-time grid, and the calibration statistic.

\begin{table}[htbp]
\centering
\caption{
    Simulation summary of empirical coverages for the six bootstrap-based methods and the Hall--Wellner band under light (L, $\sim$20\%) and heavy (H, $\sim$50\%) censoring over $[0,1]$, at the nominal 95\% level.
    The 95\% Monte Carlo reference interval, reported in percent, is $[94.865,95.135]$.
}
\label{tab:alpha005}
\small
\setlength{\tabcolsep}{4pt}
\begin{tabular}{l *{12}{c}}
% \toprule
& \multicolumn{2}{c}{$n=15$} & \multicolumn{2}{c}{$n=25$} & \multicolumn{2}{c}{$n=50$} & \multicolumn{2}{c}{$n=100$} & \multicolumn{2}{c}{$n=200$} & \multicolumn{2}{c}{$n=500$} \\
% \cmidrule(lr){2-3}\cmidrule(lr){4-5}\cmidrule(lr){6-7}\cmidrule(lr){8-9}\cmidrule(lr){10-11}\cmidrule(lr){12-13}
Method & L & H & L & H & L & H & L & H & L & H & L & H \\
% \midrule

\multicolumn{13}{@{}l}{\textit{Exponential}(1): $H_0(t)=t$} \\
Wild--Grid & $90.16$ & $87.11$ & $91.53$ & $89.31$ & $92.57$ & $91.35$ & $93.39$ & $92.57$ & $93.85$ & $93.14$ & $94.14$ & $93.86$ \\
Wild--Box & $96.26$ & $97.33$ & $96.10$ & $97.46$ & $95.76$ & $96.78$ & $95.42$ & $96.20$ & $95.23$ & $95.59$ & $94.97$ & $95.23$ \\
Weird--Grid & $86.43$ & $77.40$ & $89.48$ & $84.80$ & $91.72$ & $89.52$ & $92.94$ & $91.71$ & $93.68$ & $92.75$ & $94.06$ & $93.69$ \\
Weird--Box & $94.38$ & $95.03$ & $94.81$ & $95.75$ & $94.97$ & $95.72$ & $95.03$ & $95.53$ & $94.99$ & $95.26$ & $94.90$ & $95.08$ \\
XBoot--Grid & $83.37$ & $72.93$ & $87.70$ & $81.63$ & $90.98$ & $87.96$ & $92.62$ & $91.14$ & $93.52$ & $92.44$ & $94.01$ & $93.60$ \\
XBoot--Box & $94.03$ & $94.14$ & $94.51$ & $95.29$ & $94.75$ & $95.54$ & $94.92$ & $95.36$ & $94.93$ & $95.13$ & $94.83$ & $95.03$ \\
Hall--Wellner & $89.39$ & $84.23$ & $91.15$ & $87.16$ & $93.02$ & $90.36$ & $94.13$ & $92.28$ & $94.64$ & $93.66$ & $94.87$ & $94.45$ \\
% HW & $89.32$ & $84.07$ & $91.18$ & $87.15$ & $92.98$ & $90.20$ & $94.27$ & $92.48$ & $94.71$ & $93.60$ & $95.02$ & $94.55$ \\
% logHW & $94.13$ & $93.85$ & $94.72$ & $94.68$ & $95.12$ & $95.30$ & $95.34$ & $95.38$ & $95.22$ & $95.40$ & $95.24$ & $95.26$ \\
% \midrule

\multicolumn{13}{@{}l}{\textit{Weibull}$(0.5,0.5)$: $H_0(t)=\sqrt{2t}$} \\
Wild--Grid & $89.98$ & $83.34$ & $91.60$ & $86.84$ & $92.73$ & $90.10$ & $93.43$ & $91.65$ & $93.74$ & $92.74$ & $94.21$ & $93.49$ \\
Wild--Box & $96.14$ & $94.43$ & $96.15$ & $96.34$ & $96.20$ & $97.10$ & $95.91$ & $96.70$ & $95.35$ & $96.35$ & $95.14$ & $95.63$ \\
Weird--Grid & $81.94$ & $60.83$ & $87.87$ & $72.49$ & $91.24$ & $84.64$ & $92.74$ & $89.54$ & $93.38$ & $91.80$ & $94.09$ & $93.13$ \\
Weird--Box & $93.63$ & $90.59$ & $94.52$ & $93.62$ & $95.27$ & $95.33$ & $95.33$ & $95.75$ & $95.03$ & $95.66$ & $95.01$ & $95.28$ \\
XBoot--Grid & $77.98$ & $53.73$ & $84.94$ & $67.80$ & $89.92$ & $81.56$ & $92.23$ & $87.82$ & $93.15$ & $91.09$ & $94.02$ & $92.88$ \\
XBoot--Box & $92.45$ & $89.15$ & $93.91$ & $92.47$ & $94.80$ & $94.66$ & $95.09$ & $95.38$ & $94.90$ & $95.43$ & $94.94$ & $95.16$ \\
Hall--Wellner & $86.98$ & $78.27$ & $89.66$ & $82.44$ & $91.96$ & $87.15$ & $93.46$ & $90.33$ & $94.23$ & $92.51$ & $94.80$ & $93.95$ \\
% HW & $87.00$ & $77.89$ & $89.50$ & $82.15$ & $91.95$ & $86.99$ & $93.42$ & $90.40$ & $94.20$ & $92.46$ & $94.76$ & $94.03$ \\
% logHW & $94.14$ & $93.23$ & $94.78$ & $93.80$ & $95.10$ & $94.49$ & $95.26$ & $94.75$ & $95.22$ & $95.01$ & $95.14$ & $95.10$ \\
% \midrule

\multicolumn{13}{@{}l}{\textit{Weibull}$(2,1/\Gamma(3/2))$: $H_0(t)=\Gamma(3/2)^2 t^2$} \\
Wild--Grid & $89.87$ & $88.06$ & $91.18$ & $89.77$ & $92.44$ & $91.57$ & $93.30$ & $92.68$ & $93.77$ & $93.31$ & $94.21$ & $94.04$ \\
Wild--Box & $96.09$ & $97.74$ & $96.03$ & $97.20$ & $95.63$ & $96.47$ & $95.29$ & $95.86$ & $95.11$ & $95.37$ & $94.94$ & $95.19$ \\
Weird--Grid & $86.43$ & $82.89$ & $89.61$ & $87.06$ & $91.81$ & $90.43$ & $92.98$ & $92.10$ & $93.68$ & $93.06$ & $94.18$ & $93.95$ \\
Weird--Box & $94.53$ & $96.08$ & $94.77$ & $95.82$ & $94.89$ & $95.59$ & $94.96$ & $95.27$ & $94.92$ & $95.12$ & $94.92$ & $95.06$ \\
XBoot--Grid & $84.21$ & $79.09$ & $88.34$ & $84.80$ & $91.22$ & $89.46$ & $92.72$ & $91.72$ & $93.53$ & $92.86$ & $94.14$ & $93.84$ \\
XBoot--Box & $94.36$ & $95.68$ & $94.71$ & $95.67$ & $94.81$ & $95.49$ & $94.83$ & $95.23$ & $94.88$ & $95.05$ & $94.86$ & $95.02$ \\
Hall--Wellner & $90.25$ & $86.96$ & $91.85$ & $89.19$ & $93.49$ & $91.81$ & $94.36$ & $93.45$ & $94.76$ & $94.17$ & $94.90$ & $94.74$ \\
% HW & $90.39$ & $86.88$ & $91.83$ & $89.40$ & $93.54$ & $91.75$ & $94.43$ & $93.42$ & $94.90$ & $94.24$ & $95.05$ & $94.85$ \\
% logHW & $94.06$ & $94.10$ & $94.60$ & $94.71$ & $95.07$ & $95.40$ & $95.19$ & $95.53$ & $95.18$ & $95.37$ & $95.19$ & $95.33$ \\
% \midrule

\multicolumn{13}{@{}l}{\textit{Piecewise-constant hazard}; see \eqref{eq:piecewise_constant}} \\
Wild--Grid & $89.97$ & $88.08$ & $91.19$ & $90.01$ & $92.50$ & $91.52$ & $93.42$ & $92.84$ & $93.97$ & $93.59$ & $94.33$ & $94.28$ \\
Wild--Box & $96.14$ & $97.77$ & $96.04$ & $97.09$ & $95.62$ & $96.45$ & $95.38$ & $95.94$ & $95.23$ & $95.55$ & $95.07$ & $95.44$ \\
Weird--Grid & $86.55$ & $82.84$ & $89.64$ & $87.33$ & $91.82$ & $90.34$ & $93.08$ & $92.31$ & $93.80$ & $93.30$ & $94.28$ & $94.11$ \\
Weird--Box & $94.52$ & $96.01$ & $94.79$ & $95.79$ & $94.93$ & $95.54$ & $95.04$ & $95.41$ & $95.03$ & $95.25$ & $95.00$ & $95.28$ \\
XBoot--Grid & $84.44$ & $79.06$ & $88.35$ & $85.11$ & $91.26$ & $89.45$ & $92.78$ & $91.88$ & $93.66$ & $93.10$ & $94.24$ & $94.05$ \\
XBoot--Box & $94.38$ & $95.63$ & $94.72$ & $95.64$ & $94.80$ & $95.45$ & $94.96$ & $95.33$ & $94.98$ & $95.24$ & $94.94$ & $95.23$ \\
Hall--Wellner & $90.37$ & $86.81$ & $91.94$ & $89.34$ & $93.58$ & $91.81$ & $94.55$ & $93.44$ & $94.92$ & $94.21$ & $94.98$ & $94.91$ \\
% HW & $90.34$ & $86.88$ & $91.93$ & $89.43$ & $93.60$ & $91.79$ & $94.38$ & $93.36$ & $94.71$ & $94.12$ & $95.02$ & $94.85$ \\
% logHW & $93.76$ & $94.04$ & $94.53$ & $94.75$ & $95.02$ & $95.28$ & $95.17$ & $95.36$ & $95.08$ & $95.37$ & $95.20$ & $95.36$ \\
% \bottomrule
\end{tabular}
\end{table}

\subsection{Convergence of box-calibrated coverage}
\label{ssec:diagnostic}

\begin{figure}[ht!]
    \centering
    \includegraphics[width=\linewidth]{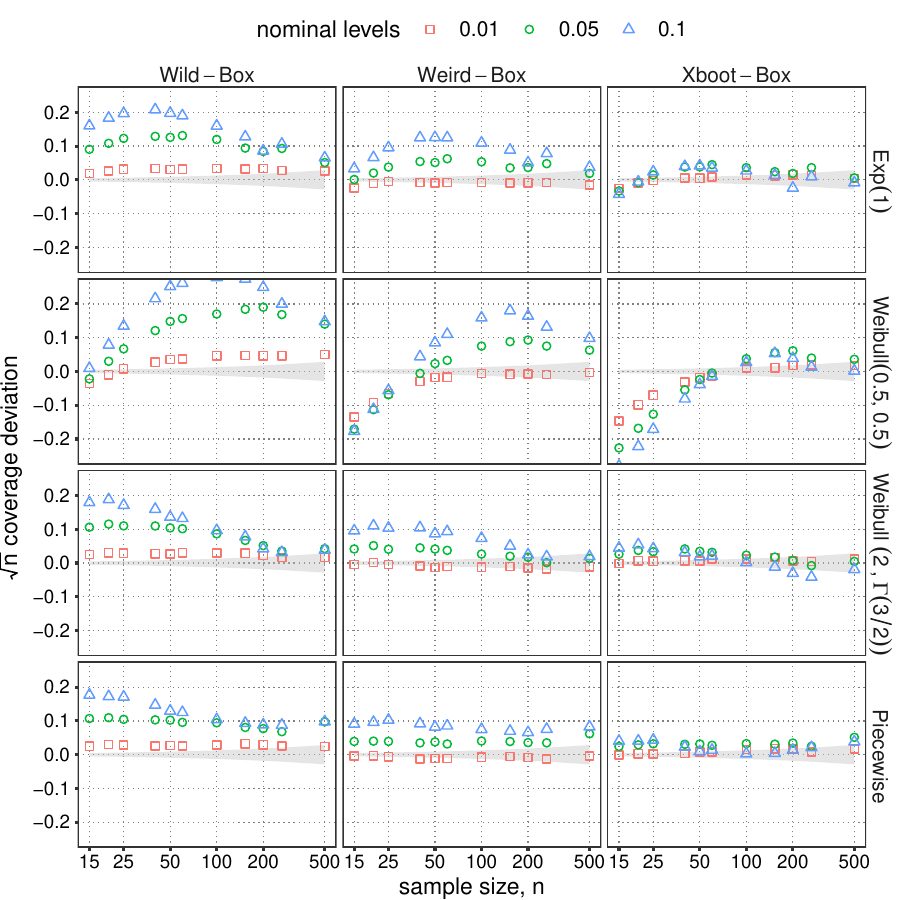}
    \caption{
        Scaled signed coverage deviation $\sqrt n\{\textsc{covg}_n(\alpha)-(1-\alpha)\}$ against sample size $n$ on a logarithmic scale, for the three box-calibrated methods under heavy censoring ($\sim 50\%$). 
        Rows correspond to the four hazard shapes; columns correspond to the three bootstrap schemes. 
        Point shape and color encode the nominal level $\alpha\in\{0.01,0.05,0.10\}$. 
        The grey ribbon marks the $95\%$ Monte Carlo reference band $\pm 1.96\sqrt{n\alpha(1-\alpha)/R}$ around zero under the nominal coverage, with $R=10^{5}$ outer replicates.
    }
    \label{fig:coverage_error_box}
\end{figure}

We compare the three box-calibrated methods under heavy censoring, where the finite-sample differences are most visible.
Figure~\ref{fig:coverage_error_box} plots the scaled signed coverage deviation $\sqrt{n}\{\textsc{covg}_n(\alpha)-(1-\alpha)\}$ against sample size; the grey ribbon is the $95\%$ Monte Carlo reference band $\pm 1.96\sqrt{n\,\alpha(1-\alpha)/R}$ around zero ($R=100{,}000$ replications), so points inside it are indistinguishable from nominal at the Monte Carlo precision.

XBoot--Box is closest to zero at every sample size and lies within or near the ribbon from moderate $n$ onward; Wild--Box stays above it, overcovering; Weird--Box lies between.
At $n=500$ under $\mathrm{Exponential}(1)$ with $\alpha=0.1$, the scaled deviations are $-0.008$, $0.037$, and $0.066$ for XBoot--, Weird--, and Wild--Box, against a ribbon half-width of $0.042$.
The exception is the decreasing-hazard $\mathrm{Weibull}(0.5,0.5)$ at the smallest sample sizes: for $n\le 25$ all three undercover and XBoot--Box is the most negative, reflecting sparse events near $\tau$; the usual ordering returns from $n\ge 50$.

A decomposition of the associated critical-value bias into a grid-distribution term and a net box-inflation term, and a grid-thinning analysis of the grid-distribution term, are given in the Supplementary Material (Sections~S4 and~S5).

\section{Melanoma data example}
\label{s:data-example}
We illustrate the proposed band using the melanoma data analyzed in
\citet{andersen1993statistical} (Example~IV.1.12). The study contains
205 patients with histological evaluation, and the data are available in
the \texttt{MASS} R package as \texttt{Melanoma}. Following the book
example, we restrict attention to the 79 male patients and use time
since operation, measured in years, as the analysis time. Death from
melanoma is treated as the event of interest, while patients who were
alive at the end of follow-up or died from other causes are treated as
censored. In this subset, 29 patients died from the disease.

\begin{figure}
    \centering
    \includegraphics[width=.8\linewidth]{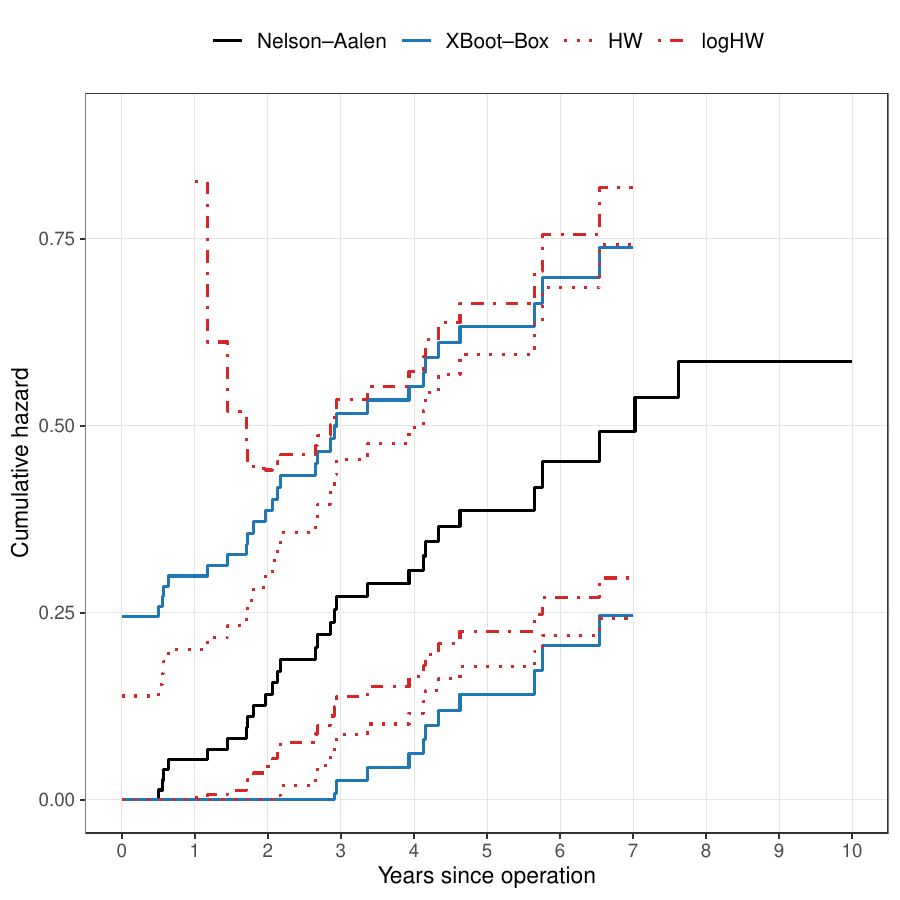}
    \caption{
    Melanoma data from \citet{andersen1993statistical}, restricted to
    the 79 male patients. The solid black step function is the
    Nelson--Aalen estimator for the cumulative hazard of death from
    melanoma. The solid blue step functions show XBoot--Box on $[0,7]$,
    the dotted red step functions show HW on $[0,7]$, and the
    dot-dashed red step functions show logHW on $[1,7]$. The logHW band
    is included to mirror the classical presentation in
    \citet{andersen1993statistical}.
    }
    \label{fig:melanoma}
\end{figure}

Figure~\ref{fig:melanoma} shows the Nelson--Aalen estimator together
with the exchangeable-bootstrap band with box calibration
(XBoot--Box), the Hall--Wellner band (HW), and the
log-transformed Hall--Wellner band (logHW). We use the interval
$[0,7]$ years for XBoot--Box and HW. For
logHW, we follow \citet{andersen1993statistical} and display it
on $[1,7]$ years, since the log transformation is not well behaved at
the origin.

Several features are visible in the figure. First, HW is the
narrowest overall. On the upper side, the log transformation moves the
HW upper bound upward over $[1,7]$, but it is unstable over
$[1,2]$. After about 3 years, the upper bound of logHW is very
close to that of XBoot--Box over roughly the interval $[3,6]$,
whereas over $[6,7]$ it becomes somewhat wider.

The lower side shows a different pattern. Over $[1,7]$,
logHW has the highest lower bound and is therefore the
narrowest in its lower half, while XBoot--Box has the lowest
lower bound and is therefore the widest. HW lies between these
two. Thus, in this example, the log transformation mainly changes
HW by widening its upper half while also pulling up its lower
half, whereas XBoot--Box gives a stable band on the full
interval $[0,7]$ and reproduces the middle-to-late upper-bound behavior
of logHW without excluding a neighborhood of zero.

\section{Discussion}
\label{s:discuss}
We proposed a procedure that combines the exchangeable bootstrap with box calibration to obtain simultaneous confidence bands for the cumulative hazard under independent right censoring.
The two components target the two aspects of the bootstrap construction introduced in Section~\ref{s:intro}:
the resampling scheme via the exchangeable bootstrap and the calibration statistic via box calibration.
% \gr{The main contribution is not a change in the first-order asymptotic limit, but rather a finite-sample correction of two distinct mismatches: a distributional mismatch arising from the resampling scheme and a geometric mismatch arising from evaluating the bootstrap discrepancy only on the observed event-time grid.} 
Both leave the first-order limit unchanged:
the exchangeable bootstrap shares the Gaussian limit of the Nelson--Aalen process, and box calibration is first-order equivalent to grid calibration. 
% \gr{The improvement therefore appears through finite-sample alignment between the bootstrap discrepancy and the oracle estimation error of the Nelson--Aalen estimator/}
% Thus the gain is not visible in the first-order weak limit, but appears in finite-samples through the way the bootstrap distribution and the calibration statistic match the original Nelson--Aalen estimator.
The improvement, therefore, appears in finite samples, through closer alignment between bootstrap discrepancy and the oracle estimation error of the Nelson--Aalen estimator.

The simulation results suggest that the resampling scheme and the calibration statistic interact in a nontrivial way.
Box calibration substantially reduces undercoverage for all three bootstrap schemes, but its effect is not simply additive across resampling methods.
When applied to wild bootstrap bands, which are already wider under grid calibration, the box adjustment can lead to overcoverage. 
By contrast, the exchangeable bootstrap preserves the ratio form of the Nelson--Aalen estimator by reweighting both the numerator and the denominator, and the corresponding box-calibrated band is usually closest to nominal coverage in the settings considered.
The ranking reversal in Section~\ref{s:sim} suggests that the exchangeable bootstrap and box calibration form the most favorable pairing in finite samples among the methods considered.

The classical route to finite-sample improvement is to construct bands on a transformed scale, typically using $g=\log$ \citep{hall1980confidence,lin1994confidence}.
A transformation can stabilize the limiting distribution and mitigate the effects of large jumps, thereby indirectly reducing the geometric mismatch addressed here.
However, the choice of transformation is not fixed by first-order theory; its performance depends on the unknown data-generating mechanism \citep{bie1987confidence,borgan1990note}, and $\log$-based bands are undefined at the origin.
Box calibration works on the original cumulative-hazard scale, requires no choice of transformation, and allows inference over the full interval $[0,\tau]$.
Pointwise studentization is another classical device, but the variance estimator can be unstable near the endpoints of $\mathcal{T}$, which, in turn, leads to boundary restrictions.
Equal-precision bands \citep{nair1984confidence} fall into this studentized class and so are not direct benchmarks for full-interval fixed-width cumulative-hazard bands on the original scale.
Box calibration could also be combined with a transformation or with studentization.

% A direct remedy might look more principled: attack the under-dispersed maximum at its source with high-dimensional central-limit machinery \citep{chernozhukov2017central,koike2026high}.
% But any such correction must estimate the high-dimensional dependence among the $m_n$ increments, the very object whose maximum the bootstrap under-disperses.
% Box calibration estimates nothing.
% It leaves the bootstrap maximum untouched and adds an inflation of scale $1/R_{nj^\ast}$, already present in the estimator and fixed by the data, and reuses the step values formed for grid calibration, so the two calibrations have the same cost.
The grid statistic can also be viewed through the lens of high-dimensional maximum approximation \citep{chernozhukov2017central,koike2026high}. 
It is the maximum over $m_n$ event-time coordinates. 
A correction based directly on this viewpoint would need to approximate the dependence structure across the $m_n$ event-time coordinates, which is itself random through the observed event grid and risk sets. 
Box calibration takes a different route. 
It leaves the bootstrap law on the observed grid unchanged, and adds the local correction using jump values already computed for grid calibration. 
Thus the correction keeps the computational cost of grid calibration while accounting for a finite-sample feature that the grid maximum alone does not record.

% The bands here are fixed-width and two-sided, with a single critical value. Calibrating the upper and lower deviations separately could exploit the asymmetry of the geometric mismatch, since the upper side involves the right endpoint of each inter-event interval and the lower side the jump point.

The step-function ratio structure also appears in several related estimators. Natural extensions include the Kaplan--Meier product-integral estimator, the Aalen--Johansen estimator for competing risks and multistate models \citep{aalen1978empirical}, the Breslow baseline cumulative hazard in the Cox model \citep{cox1972regression,breslow1975analysis}, and settings with left truncation. 
For multivariate monotone step-function estimators on random grids, an analogous calibration would replace each grid cell by a higher-dimensional box.
In each case the bootstrap and original estimators share the same jump grid, the feature behind the first-order equivalence of grid and box calibration.
\citet{dietrich2025wild} note that Efron's bootstrap appears to resample the general Nelson--Aalen estimator correctly when individuals can experience multiple events.
Since Efron's bootstrap is a special case of the exchangeable bootstrap, the empirical-process argument used here should extend to that setting under conditions analogous to Condition~\ref{cond:1}.

% Throughout we assume an absolutely continuous $H_0$, so that event times are distinct almost surely. Real data contain ties, which merge jumps and enlarge the inter-jump intervals, and so widen the very between-jump gap that box calibration brackets. A formal treatment of ties, including how the box statistic should apportion a tied jump, is left to future work.

% Finally, box calibration can be viewed as one member of a larger family.
% For $\lambda\in [0,1]$, a $\lambda$-box statistic would use only a fraction $\lambda$ of each bootstrap jump, with $\lambda=0$ recovering grid calibration and $\lambda=1$ giving the full box.
% All choices of $\lambda$ are expected to share the same first-order limit, but their finite-sample behavior may differ.
% This may be especially relevant when tied event times are present.
% A tied jump represents multiple events recorded at the same time; treating the entire tied jump as a single vertical box may be too conservative, while using only part of the tied jump may better reflect the fact that the tie is often a consequence of time discretization.
% Developing data-driven or tie-aware choices of $\lambda$ is therefore a useful direction for future work.

\section*{Declaration of the use of generative AI and AI-assisted technologies}

During the preparation of this work the authors used Claude Opus 4.8 and ChatGPT 5.5 for broad literature search, grammar check, and package documentation. After using these tools the authors reviewed and edited the content as necessary and take full responsibility for the content of the publication.

% \section*{Acknowledgement}
% Acknowledgements should appear after the body of the paper but before any appendices and be as brief as possible
% subject to politeness. 
%Information, such as contract numbers, of no interest to readers, should be excluded.

\section*{Supplementary material}
\label{SM}
The Supplementary Material provides all technical details and proofs, as well as additional numerical simulations.

%Further instructions will be given when a paper is accepted.
%\vspace*{-10pt}

\bibliographystyle{biometrika}
\bibliography{boxcalibration.bib}

\end{document}

\nolinenumbers
\jname{Biometrika}
%% The year, volume, and number are determined on publication
\jyear{2025}
\jvol{112}
\jnum{1}
\cyear{2025}
%% The \doi{...} and \accessdate commands are used by the production team
%\doi{10.1093/biomet/asm023}
\accessdate{Advance Access publication on 14 February 2025}

%% These dates are usually set by the production team
\received{d M yyyy}
\revised{d M yyyy}

%% The left and right page headers are defined here:
\markboth{M. Lin, G. Rempala, E. Kenah, and Q. Lin}{Exchangeable bootstrap and box calibration}

%% Here are the title, author names and addresses
\title{Supplementary material for Simultaneous confidence bands for cumulative hazard via exchangeable bootstrap and box calibration} % all lower case

% , , Columbus, Ohio, U.S.A. 
\author{Min Lin}
\affil{Division of Biostatistics, College of Public Health, The Ohio State University,\\ Columbus, Ohio 43210, U.S.A.
\email{lin.5267@osu.edu}}

\author{Grzegorz Rempala}
\affil{Division of Biostatistics, College of Public Health, The Ohio State University,\\ Columbus, Ohio 43210, U.S.A.
\email{rempala.3@osu.edu}}

\author{Eben Kenah}
\affil{Division of Biostatistics, College of Public Health, The Ohio State University,\\ Columbus, Ohio 43210, U.S.A.
\email{kenah.1@osu.edu}}

\author{\and Qianying Lin}
\affil{Division of Biostatistics, College of Public Health, The Ohio State University,\\ Columbus, Ohio 43210, U.S.A. 
\email{lin.5268@osu.edu}}

\maketitle

\section{Weak convergence framework}

For completeness, we record the weak-convergence notions used in the proofs; see Chapter~1 of \citet{van_der_vaart_weak_2023} for a full treatment. Let $(\mathbb D,d)$ be a metric space, let $C_b(\mathbb D)$ denote the class of bounded, continuous functions $f\colon \mathbb D\to\mathbb R$, and let $X_n\colon \Omega_n\to\mathbb D$ be arbitrary, possibly nonmeasurable, maps defined on probability spaces $(\Omega_n,\mathscr F_n,P_n)$.

For a possibly nonmeasurable real-valued map $T$, define the outer expectation
\[
\E^* T \coloneqq \inf\bigl\{\E U \colon U\ge T,\ U \text{ measurable, and } \E U \text{ exists}\bigr\},
\]
and the outer probability by
\[
\P^*(A)\coloneqq \E^*(\mathbbm 1_A).
\]
If $\E^*T<\infty$, there exists a minimal measurable majorant $T^*$ such that $T\le T^*$ and $\E^*T=\E T^*$.

If $X$ is a Borel measurable $\mathbb D$-valued random element, we write $X_n\rightsquigarrow X$ if
\[
\E^* f(X_n)\to \E f(X),
\qquad \text{for every } f\in C_b(\mathbb D).
\]
We say that $X$ has separable Borel law if its law is concentrated on a separable Borel subset of $\mathbb D$.

Let
\[
BL_1(\mathbb D)\coloneqq
\bigl\{
f\colon \mathbb D\to\mathbb R \colon
\sup_{x\in\mathbb D}|f(x)|\le 1,\ 
\sup_{x\ne y}|f(x)-f(y)|/d(x,y)\le 1
\bigr\}.
\]
If $X$ has separable Borel law, then
\[
X_n\rightsquigarrow X
\quad\Longleftrightarrow\quad
\sup_{f\in BL_1(\mathbb D)}
\bigl|
\E^* f(X_n)-\E f(X)
\bigr|
\to 0.
\]
This bounded-Lipschitz formulation is the one used throughout the supplement.

A sequence $X_n$ converges to $X$ in outer probability, denoted
$X_n\overset{\P^*}{\to}X$, if
\[
\P^*\bigl(d(X_n,X)>\varepsilon\bigr)
\equiv
\P\bigl(d(X_n,X)^*>\varepsilon\bigr)\to 0
\qquad \text{for every } \varepsilon>0.
\]
If $d(X_n,X)$ is measurable for each $n$, then
$X_n\overset{\P^*}{\to}X$ is equivalent to $X_n\overset{\P}{\to}X$.

Next, let $W_n$ be a random vector in $\mathbb R^n$, and let
$\phi_n\colon \mathbb D\times \mathbb R^n\to \mathbb D$ satisfy that
$w\mapsto\phi_n(x,w)$ is Borel measurable for every fixed
$x\in\mathbb D$ and every $n$. When $X$ has separable Borel law, we say
that $\phi_n(X_n,W_n)$ converges conditionally weakly to $X$ in outer
probability, denoted
$\phi_n(X_n,W_n)\overunderset{\P}{W}{\rightsquigarrow}X$, if
\[
\sup_{f\in BL_1(\mathbb D)}
\bigl|
\E_{W_n}f\bigl(\phi_n(X_n,W_n)\bigr)-\E f(X)
\bigr|
\overset{\P^*}{\longrightarrow}0
\]
and, for every $f\in BL_1(\mathbb D)$,
\[
\E_{W_n}\bigl[f\bigl(\phi_n(X_n,W_n)\bigr)^*\bigr]
-
\E_{W_n}\bigl[f\bigl(\phi_n(X_n,W_n)\bigr)_*\bigr]
\overset{\P^*}{\longrightarrow}0,
\]
where $\E_{W_n}$ denotes the section integral over the $W_n$-coordinate, with $X_n$ held fixed, $f(\phi_n(X_n,\\W_n))^*$ and $f(\phi_n(X_n,W_n))_*$ denote measurable majorants and minorants of $f(\phi_n(X_n,W_n))$ with respect to $(X_n,W_n)$ jointly, and $Z_*\coloneqq -(-Z)^*$.

All subsequent uses of weak convergence, conditional weak convergence, and the associated continuous mapping and functional delta arguments are understood in this sense.
\section{Proofs}
\subsection{Proof of Theorem~1}
\begin{proof}
Let $Z_i\coloneqq T_i\land C_i$, $\delta_i\coloneqq \mathbbm 1\{T_i\le C_i\}$, and $X_i\coloneqq (Z_i,\delta_i)\in\mathfrak X\coloneqq \mathbb R\times \{0,1\}$. Suppose $X_i$ are i.i.d.\ with law $P$. Let $\pi_1(z,\delta)=z$ and $\pi_2(z,\delta)=\delta$ be the coordinate projections on $\mathfrak X$. Let $\tau>0$ be given and define $\mathcal T\coloneqq [0,\tau]$. 
\begin{itemize}
    \item $\mathcal G_1\coloneqq\{\mathbb R\to\mathbb [0,1]\colon z\mapsto \mathbbm 1_{[0,t]}(z)\colon t\in\mathcal T\}$ and $\mathcal G_2\coloneqq \{\mathbb R\to[0,1]\colon z\mapsto \mathbbm 1_{[t,\infty)}(z)\colon t\in\mathcal T\}$ are sets of monotone and uniformly bounded functions on $\mathbb R$. By the bracketing entropy bound (Theorem~2.7.9 of VW) and the bracketing Donsker theorem (Theorem~2.5.6 of VW), $\mathcal G_1$ and $\mathcal G_2$ are $P\circ \pi_1^{-1}$-Donsker on $\mathbb R$. It follows that $\mathcal G_1\circ \pi_1\coloneqq \{g\circ \pi_1\colon g\in\mathcal G_1\}$ and $\mathcal G_2\circ \pi_1\coloneqq \{g\circ \pi_1\colon g\in\mathcal G_2\}$ are $P$-Donsker on $\mathfrak X$.
    \item $\mathcal G_3\coloneqq \{\{0,1\}\to\{0,1\}\colon\delta\mapsto\delta\}$ a singleton class is trivially $P\circ \pi_2^{-1}$-Donsker, so $\mathcal G_3\circ \pi_2$ is $P$-Donsker on $\mathfrak X$.
    \item Since $\mathcal G_1\circ \pi_1$ and $\mathcal G_3\circ \pi_2$ are uniformly bounded $P$-Donsker classes, the pairwise products $\mathcal F_1\coloneqq (\mathcal G_1\circ \pi_1)\cdot (\mathcal G_3\circ \pi_2)=\{\mathfrak X\to \mathbb R\colon(z,\delta)\mapsto \delta \mathbbm 1_{[0,t]}(z)\colon t\in\mathcal T\}$ is $P$-Donsker (Example~2.10.10 of VW). Denote $\mathcal F_2\coloneqq \mathcal G_2\circ \pi_1=\{\mathfrak X\to\mathbb R\colon (z,\delta)\mapsto \mathbbm 1_{[t,\infty)}(z)\colon t\in\mathcal T\}$. Then, $\mathcal F\coloneqq \mathcal F_1\cup \mathcal F_2$ is $P$-Donsker as well (Example~2.10.9 of VW). Note that $\mathcal F_1 \cap \mathcal F_2=\varnothing$ since every function in $\mathcal F_1$ depends on $\delta$ while every function in $\mathcal F_2$ is not.
    % \qyl{Note that $\mathcal F_1 \cap \mathcal F_2=\varnothing$ since every function in $\mathcal F_1$ depends on $\delta$ while every function in $\mathcal F_2$ is not.}
    \item $\mathcal G_1'\coloneqq \{\mathbb R\to[0,1]\colon z\mapsto \mathbbm 1_{[0,q]}(z)\colon q\in(\mathbb Q\cap \mathcal T)\cup\{\tau\}\}$ is a countable subset of $\mathcal G_1$ such that for every $g\in\mathcal G_1$, there exists a sequence $\{g_m\}$ in $\mathcal G_1'$ with $g_m(z)\to g(z)$ for every $z\in\mathbb R$. We say $\mathcal G_1$ is pointwise measurable (PM). Similarly, $\mathcal G_2$ is PM by $\mathcal G_2'\coloneqq \{\mathbb R\to[0,1]\colon z\mapsto \mathbbm 1_{[q,\infty)}\colon q\in(\mathbb Q\cap \mathcal T)\cup\{\tau\}\}$. The singleton $\mathcal G_3$ is trivially PM. Coordinate projections, pairwise products, and finite union do not change the PM property. So $\mathcal F$ is PM with envelope $F(z,\delta)\coloneqq \sup_{f\in\mathcal F}|f(z,\delta)|$ uniformly bounded by $1$. By Proposition~8.11 of \citet{kosorok_introduction_2008}, $\mathcal F_\eta\coloneqq \{f_1-f_2\colon f_1,f_2\in\mathcal F,\int |f_1-f_2|^2\,\mathrm dP<\eta^2\}$ is PM for every $\eta>0$.
\end{itemize}
Define the empirical process $\mathbb G_n$ as a map into $\ell^\infty(\mathcal F)$ by 
\begin{equation*}
    \mathbb G_nf=\sqrt{n}\left(\frac1n\sum_{i=1}^nf(X_i)-\int f\,\mathrm{d}P\right).
\end{equation*}
Since $\mathcal F$ is $P$-Donsker,
\begin{equation*}
    \mathbb G_n\rightsquigarrow\mathbb G\qquad \text{in }\ell^\infty(\mathcal F)
\end{equation*}
where $\mathbb G$ is a tight zero-mean Gaussian process in $\ell^\infty(\mathcal F)$ with covariance structure
\begin{equation*}
    \E \mathbb Gf_1\mathbb G f_2=\int f_1f_2\,\mathrm dP-\int f_1\,\mathrm dP\int f_2\,\mathrm dP.
\end{equation*}
Let $W_{n1},\ldots,W_{nn}$ be weights satisfying Condition 1. Define weighted bootstrap empirical process $\widehat{\mathbb G}^W_n$ by
\begin{equation*}
    \widehat{\mathbb G}^W_nf=\frac1{\sqrt n}\sum_{i=1}^n(W_{ni}-1)f(X_i).
\end{equation*}
By exchangeable bootstrap central limit theorem (Theorem~3.7.13 of VW), 
\begin{equation*}
    \widehat{\mathbb G}^W_n\overunderset{\P}{W}{\rightsquigarrow}\mathbb G\qquad \text{in }\ell^\infty(\mathcal F).
\end{equation*}
\textit{From $\ell^\infty(\mathcal F)$ to $\ell^\infty(\mathcal T)^2$.} Recall that $\mathcal F=\mathcal F_1\cup \mathcal F_2$ and $\mathcal F_1\cap \mathcal F_2=\varnothing$. To separate functions, we denote $f_{t,1}$ as a function in $\mathcal F_1$ and $f_{t,2}$ as a function in $\mathcal F_2$. Define map $J\colon \ell^\infty(\mathcal F)\to \ell^\infty(\mathcal T)^2$ by
\[
J[G](t)\coloneqq \big(G(f_{t,1}),G(f_{t,2})\big),\qquad t\in\mathcal T.
\]
$J$ is linear: for $a,b\in\mathbb R$ and $G_1,G_2\in \ell^\infty(\mathcal F)$, 
\begin{align*}
    J[aG_1+bG_2](t)&=\Big((aG_1+bG_2)(f_{t,1}),(aG_1+bG_2)(f_{t,2})\Big)\\
    &=\Big(aG_1(f_{t,1})+bG_2(f_{t,1}),aG_1(f_{t,2})+bG_2(f_{t,2})\Big)\\
    &=aJ[G_1](t)+bJ[G_2](t).
\end{align*}
$J$ is isometry: for $G\in\ell^\infty(\mathcal F)$
\begin{equation*}
    \|J[G]\|_{\ell^\infty(\mathcal T)^2}=\max_{j\in\{1,2\}}\sup_{t\in\mathcal T}|G(f_{t,j})|
=\sup_{f\in\mathcal F}|G(f)|=\|G\|_{\mathcal F}.
\end{equation*}
Therefore, for any $f\in BL_1(\ell^\infty(\mathcal T))$, $f\circ J\in BL_1(\ell^\infty(\mathcal F))$. It follows from continuous mapping that
\begin{align*}
    J[\mathbb G_n]=\sqrt n(\bar N_n-\E N,\bar Y_n-\E Y)&\rightsquigarrow J[\mathbb G]\qquad \text{in }\ell^\infty(\mathcal T)^2\\
    J[\widehat{\mathbb G}^W_n]=\sqrt n(\bar N^W_n- \bar N_n,\bar Y^W_n- \bar Y_n)&\overunderset{\P}{W}{\rightsquigarrow}J[\mathbb G]\qquad \text{in }\ell^\infty(\mathcal T)^2
\end{align*}
where $J[\mathbb G]$ is still a tight zero-mean Gaussian process by Lemma~3.10.8 of VW. Denote $J[\mathbb G]=(\mathbb G_N,\mathbb G_Y)$. For given $M,\varepsilon>0$, the map $(A,B)\mapsto \int_0^\cdot \frac{\mathrm d A}{B}$ is Hadamard-differentiable on the domain $\{(A,B)\colon \int |\mathrm dA|\le M,B\ge \varepsilon\}$ at every point $(A,B)$ such that $1/B$ is of bounded variation \citep{gill_survey_1990}. By the assumption that $\inf_{t\in\mathcal T}\E Y(t)>0$, $(\bar N_n,\bar Y_n)$ and $(\bar N_n^W,\bar Y_n^W)$ are contained in this domain with probability tending to $1$ for $M\ge 1$ and sufficiently small $\varepsilon$. The functional delta-method (Theorem~3.10.4 of VW) and conditional functional delta-method (Theorem~3.10.11 of VW) imply
\begin{align*}
    \sqrt{n}\left(\widehat H_n-H_0\right)=\sqrt{n}\int_0^\cdot \frac{\text{d}\bar N_n-\bar Y_n\,\mathrm{d}H_0}{\E Y}+o_{\P}^*(1)&\rightsquigarrow \mathbb Z\qquad \text{in }\ell^\infty(\mathcal T)\\
    \sqrt{n}\left(\widehat H^W_n-\widehat H_n\right)&\overunderset{\P}{W}{\rightsquigarrow} \mathbb Z\qquad \text{in }\ell^\infty(\mathcal T)
\end{align*}
where $\mathbb Z$ is a tight zero-mean Gaussian process with covariance structure $\E \mathbb Z(s)\mathbb Z(t)=\int_0^{s\land t}\{\E Y(u)\}^{-1}\,\mathrm d H_0(u)$. 
\end{proof}

\subsection{Proof of Proposition~1}
\begin{proof}
    We have $\sqrt{n}(\widehat H^W_n-\widehat H_n)\overunderset{\P}{W}{\rightsquigarrow} \mathbb Z$ by Theorem~1. Note that $g\colon \ell^\infty(\mathcal T)\to \mathbb R\colon G\mapsto \|G\|_{\mathcal T}$ is $1$-Lipschitz: $\big|g(G_1)-g(G_2)\big|\le \|G_1-G_2\|_{\mathcal T}$ by the triangle inequality. Therefore, for every $f\in BL_1(\mathbb R)$, $f\circ g\in BL_1(\ell^\infty(\mathcal T))$. It follows from definition that 
    \begin{equation*}
        \sqrt n\widehat \Delta_{n,\mathrm{grid}}^W=g\!\left(\sqrt{n}(\widehat H^W_n-\widehat H_n)\right) \overunderset{\P}{W}{\rightsquigarrow}g (\mathbb Z)=\|\mathbb Z\|_{\mathcal T}
    \end{equation*}
    Also by continuous mapping,
    \begin{equation*}
        \sqrt n\Delta_n=g\!\left(\sqrt n(\widehat H_n-H_0)\right)\rightsquigarrow g(\mathbb Z)=\|\mathbb Z\|_{\mathcal T}.
    \end{equation*}
    Let $F(x)\coloneqq \P(\|\mathbb Z\|_{\mathcal T}\le x)$ for $x\in\mathbb R$. By Lemma~1.5.9 of VW, $\mathbb Z$ is uniformly continuous under its intrinsic covariance semimetric
    \begin{align*}
        \rho_2(s,t)\coloneqq \sqrt{\E|\mathbb Z(s)-\mathbb Z(t)|^2}=\sqrt{\int_{s\land t}^{s\lor t}\frac{\mathrm dH_0(u)}{\E Y(u)}}
    \end{align*}
    and $(\mathcal T,\rho_2)$ is a totally bounded metric space. Therefore $\|\mathbb Z\|_{\mathcal T}=\|\mathbb Z\|_{\mathcal T_0}$ almost surely for a countable subset $\mathcal T_0\subseteq \mathcal T$. It follows that $F$ is continuous everywhere and strictly increasing on $[0,\infty)$ \citep{gaenssler_continuity_2007}. Denote $\widehat F_n(x)\coloneqq \P\bigl(\sqrt n\widehat \Delta_{n,\mathrm{grid}}^W\le x\mid \mathcal D_n\bigr)$ and $F_n(x)\coloneqq \P\bigl(\sqrt n\Delta_n\le x\bigr)$. 
    
    By Lemmas~10.11 and 10.12 of \citet{kosorok_introduction_2008}, 
    \begin{align*}
        \sup_{x\in\mathbb R}\left|\widehat F_n(x)-F(x)\right|&\overset{\P}{\to}0,\quad         \sup_{x\in\mathbb R}\left|F_n(x)-F(x)\right|\to0.
    \end{align*}
    It follows from the triangle inequality and rescaling that 
    \begin{equation*}
        \sup_{x\in\mathbb R}\left|\P\bigl(\widehat \Delta_{n,\mathrm{grid}}^W\le x\mid \mathcal D_n\bigr)-\P\bigl(\Delta_n\le x\bigr)\right|=\sup_{x\in\mathbb R}\left|\widehat F_n(\sqrt{n}x)-F_n(\sqrt{n}x)\right|=\sup_{x\in\mathbb R}\left|\widehat F_n(x)-F_n(x)\right|\overset{\P}{\to}0.
    \end{equation*}
    Now let $\alpha\in(0,1)$ be given and define
    \begin{equation*}
        \hat q_{n,1-\alpha}\coloneqq \inf\{x\colon \widehat F_n(x)\ge 1-\alpha\},\quad q_{1-\alpha}\coloneqq \inf \{x\colon F(x)\ge 1-\alpha\}>0.
    \end{equation*}
    Let $\varepsilon>0$ be given and take
    \begin{equation*}
        \eta_{\varepsilon}\coloneqq \min \left\{(1-\alpha)-F(q_{1-\alpha}-\varepsilon),\,F(q_{1-\alpha}+\varepsilon)-(1-\alpha)\right\}.
    \end{equation*}
    Since $F$ is continuous and strictly increasing at $q_{1-\alpha}$, we have $\eta_{\varepsilon}>0$. On the event $\sup_{x\in\mathbb R}|\widehat F_n(x)-F(x)|<\eta_{\varepsilon}$, we get $\widehat F_n(q_{1-\alpha}-\varepsilon)<1-\alpha$ and $\widehat F_n(q_{1-\alpha}+\varepsilon)>1-\alpha$, hence by definition of the quantile,
    \begin{equation*}
        q_{1-\alpha}-\varepsilon<\hat q_{n,1-\alpha}\le q_{1-\alpha}+\varepsilon.
    \end{equation*}
    Therefore,
    \begin{equation*}
        \P(|\hat q_{n,1-\alpha}-q_{1-\alpha}|>\varepsilon)\le \P(\sup_{x\in\mathbb R}|\widehat F_n(x)-F(x)|\ge\eta_{\varepsilon})=o(1).
    \end{equation*}
    Since
    \begin{align*}
        &\{\sqrt n\Delta_n\le q_{1-\alpha}-\varepsilon\}\cap \{|\hat q_{n,1-\alpha}-q_{1-\alpha}|\le \varepsilon\}\\
        \subset\,& \{\sqrt n\Delta_n\le \hat q_{n,1-\alpha}\}\\
        \subset\,&\{\sqrt n\Delta_n\le q_{1-\alpha}+\varepsilon\}\cup \{|\hat q_{n,1-\alpha}-q_{1-\alpha}|> \varepsilon\},
    \end{align*}
    taking probability gives
    \begin{align*}
        F_n(q_{1-\alpha}-\varepsilon)-\P(|\hat q_{n,1-\alpha}-q_{1-\alpha}|> \varepsilon)&\le \P(\sqrt n\Delta_n\le \hat q_{n,1-\alpha})\\
        &\le F_n(q_{1-\alpha}+\varepsilon)+\P(|\hat q_{n,1-\alpha}-q_{1-\alpha}|> \varepsilon).
    \end{align*}
    Letting $n\to \infty$ we obtain
    \begin{align*}
        F(q_{1-\alpha}-\varepsilon)\le \liminf_n \P(\sqrt n\Delta_n\le \hat q_{n,1-\alpha})\le \limsup_n\P(\sqrt n\Delta_n\le \hat q_{n,1-\alpha})\le F(q_{1-\alpha}+\varepsilon). 
    \end{align*}
    Now let $\varepsilon\downarrow 0$ and use continuity of $F$ at $q_{1-\alpha}$ to get $\P(\sqrt n\Delta_n\le \hat q_{n,1-\alpha})\to F(q_{1-\alpha})=1-\alpha$. Finally, since $\hat q_{n,1-\alpha}=\sqrt{n}\hat q_{n,\mathrm{grid}}^W(1-\alpha)$, $\P(\Delta_n\le \hat q_{n,\mathrm{grid}}^W(1-\alpha))\to 1-\alpha$.
\end{proof}

\subsection{Proof of Theorem~2}
\begin{proof}
    Let $Z_i\coloneqq T_i\land C_i$, $\delta_i\coloneqq \mathbbm 1\{T_i\le C_i\}$, and $X_i\coloneqq (Z_i,\delta_i)\in\mathfrak X\coloneqq \mathbb R\times \{0,1\}$. Work on the product probability space
    \begin{equation*}
        (\Omega,\mathscr F,\P)=(\mathfrak X^\infty\times \mathfrak W,\mathscr X^\infty\otimes \mathscr W,P^\infty\times \P_W)    
    \end{equation*}
    where $X_i$ is the $i$th coordinate projection on $\mathfrak X^\infty$ and the exchangeable bootstrap weights depend only on the $\mathfrak W$-coordinate. Hence, $X_1,X_2,\ldots$ are i.i.d.\ with law $P$, and the weights are independent from the data. Write $\P_X\coloneqq P^\infty$, and let $\E_W$ denote the section integral over the $\mathfrak W$-coordinate. Whenever minimal measurable majorants are used, they are taken with respect to the joint $(\mathscr X^\infty\otimes \mathscr W)$-measurable structure. Let $\widetilde{H}_n^W$ be any monotone continuous interpolation of $\widehat H_n^W$. 

By the triangle inequality,
    \begin{align*}
        \sup_{f\in BL_1(\ell^\infty(\mathcal T))}\left|\E_Wf(\sqrt n(\widetilde H_n^W-\widehat H_n))-\E f(\mathbb Z)\right|\le I_n+II_n 
    \end{align*}
    where 
    \begin{align*}
        I_n&=\sup_{f\in BL_1(\ell^\infty(\mathcal T))}\left|\E_Wf(\sqrt n(\widehat H_n^W-\widehat H_n))-\E f(\mathbb Z)\right|\overset{\P^*_X}{\longrightarrow}0
    \end{align*}
    by Theorem~1, and
    \begin{align*}
        II_n&=\sup_{f\in BL_1(\ell^\infty(\mathcal T))}|\E_Wf(\sqrt n(\widetilde H_n^W-\widehat H_n))-\E_W f(\sqrt n(\widehat H_n^W-\widehat H_n))|\\
        &\le \E_W\{\|\sqrt n(\widetilde H_n^W-\widehat H^W_n)\|^*_{\mathcal T}\land 2\}\\
        &\le \varepsilon+2\P_W(\|\sqrt n(\widetilde H_n^W-\widehat H^W_n)\|^*_{\mathcal T}>\varepsilon)\eqqcolon \varepsilon+2R_n(\varepsilon)
    \end{align*}
    for every $\varepsilon>0$. Since $R_n(\varepsilon)$ is measurable, it is enough to prove that $R_n(\varepsilon)\to 0$ in $\P_X$-probability for every $\varepsilon>0$. Once this has been established, we obtain $\sqrt n(\widetilde H_n^W-\widehat H_n)\overunderset{\P}{W}{\rightsquigarrow}\mathbb Z$, and the same argument as Proposition~1 gives the desired result.

    By the strong law of large numbers, $\bar Y_n(\tau)\to y_\tau\coloneqq \E Y(\tau)>0$ $\P_X$-almost surely. Since $H_0$ is absolutely continuous, the observed event times are distinct $\P_X$-almost surely for every $n$. Let $\mathfrak X_0^\infty$ be the intersection of these almost-sure events, so that $\P_X(\mathfrak X_0^\infty)=1$. For each $x=(x^1,x^2,\ldots)\in\mathfrak X_0^\infty$, write $x_n=(x^1,\ldots,x^n)$ and define 
\begin{equation*}
    0=E_{n,0}(x_n)<E_{n,1}(x_n)<\cdots<E_{n,m_n(x_n)}(x_n)<\tau<E_{n,m_n(x_n)+1}(x_n),
\end{equation*}
where $E_{n,1}(x_n),\ldots,E_{n,m_n(x_n)}(x_n)$ are the discontinuity points of $t\mapsto \widehat H_n(x_n,t)$ on $[0,\tau]$, and $E_{n,m_n(x_n)+1}(x_n)$ is the next discontinuity point after $\tau$, possibly equal to $\infty$. When $E_{n,m_n(x_n)+1}(x_n)=\infty$, there is no uncensored event after $\tau$, and $K_{n,\tau}^W(x_n)=\widehat H_n^W(E_{n,m_n(x_n)}(x_n))$. Outside $\mathfrak X_0^\infty$, define these quantities arbitrarily. We suppress the argument $x_n$ whenever there is no ambiguity.

Fix $\varepsilon>0$. By construction
\begin{align*}
    \|\sqrt n(\widetilde H_n^W-\widehat H_n^W)\|^*_{\mathcal T}=(A_n^W\vee B_n^W)^*=(A_n^W)^*\vee (B_n^W)^*
\end{align*}
where 
\begin{align*}
    A_n^W&\coloneqq \sqrt n\max_{i=1,\ldots,m_n}\left\{\widehat H_n^W(E_{n,i})-\widehat H_n^W(E_{n,i-1})\right\},\\
    % B_n^W&\coloneqq \sqrt n\frac{(\tau -E_{n,m_n})\{\widehat H_n^W(E_{n,m_n+1})-\widehat H_n^W(E_{n,m_n})\}}{E_{n,m_n+1}-E_{n,m_n}}.
    B_n^W&\coloneqq \sqrt n \left\{K_{n,\tau}^W-\widehat H_n^W(E_{n,m_n})\right\}\in\left[0,\sqrt n\left\{\widehat H_n^W(E_{n,m_n+1})-\widehat H_n^W(E_{n,m_n})\right\}\right].
\end{align*}
Hence,
\begin{align*}
    R_n(\varepsilon)\le \P_W((A_n^W)^*>\varepsilon)+\P_W((B_n^W)^*>\varepsilon).
\end{align*}
We first treat $A_n^W$. Let $\pi(j)$ be the subscript of the $j$th uncensored event. Then
\begin{equation*}
    A_n^W = \max_{j=1,\ldots,m_n} \frac{W_{n,\pi(j)}}{\sqrt n\,\bar Y_n^W(E_{n,j})} \le \frac{\max_{i=1,\ldots,n}W_{ni}/\sqrt n}{\bar Y_n^W(\tau)},
\end{equation*}
where $\bar Y_n^W(\tau)=n^{-1}\sum_{i=1}^nW_{ni}\mathbbm 1\{Z_i\ge \tau\}$ is $(\mathscr X^\infty\otimes \mathscr W)$-measurable. Therefore,
\begin{align*}
    \P_W((A_n^W)^*>\varepsilon)&\le \P_W((A_n^W)^*>\varepsilon,\bar Y_n^W(\tau)\ge y_\tau/2)+\P_W((A_n^W)^*>\varepsilon,\bar Y_n^W(\tau)< y_\tau/2)\\
    &\le \P_W\!\left(\max_{1\le i\le n}W_{ni}/\sqrt n>\varepsilon y_\tau/2\right) + \P_W\!\left(\bar Y_n^W(\tau)< y_\tau/2\right)
\end{align*}
where the first term is bounded above by 
\begin{align*}
    \frac{2}{\varepsilon y_\tau}\E_W \frac{\max_{1\le i\le n}|W_{ni}-1|+1}{\sqrt n}=o(1)\qquad \text{using Condition~1}.
\end{align*}
For the second term, the proof of Theorem~1 gives $\sqrt n(\bar Y_n^W-\bar Y_n)\overunderset{\P}{W}{\rightsquigarrow}\mathbb G_Y$ in $\ell^\infty(\mathcal T)$ for a tight zero-mean Gaussian process $\mathbb G_Y$. The evaluation map $\ell^\infty(\mathcal T)\to \mathbb R\colon G\mapsto G(\tau)$ is linear and $1$-Lipschitz, whence
\begin{equation*}
    U_n\coloneqq \sqrt n(\bar Y_n^W(\tau)-\bar Y_n(\tau))\overunderset{\P}{W}{\rightsquigarrow}\mathbb G_Y(\tau).
\end{equation*}
Set $V_n\coloneqq \bar Y_n^W(\tau)-\bar Y_n(\tau)= U_n/\sqrt n$. For each $f\in BL_1(\mathbb R)$, define $g_{n,f}(x)\coloneqq f(x/\sqrt n)$. Then $g_{n,f}\in BL_1(\mathbb R)$ as well and $\E_Wf(V_n)=\E_Wg_{n,f}(U_n)$. Hence,
\begin{align*}
    \sup_{f\in BL_1(\mathbb R)}\left|\E_Wf(V_n)-f(0)\right|&\le \sup_{f\in BL_1(\mathbb R)}\left|\E_Wg_{n,f}(U_n)-\E g_{n,f}(\mathbb G_Y(\tau))\right|\\
    &\qquad +\sup_{f\in BL_1(\mathbb R)}\left|\E f\!\Big(\frac{\mathbb G_Y(\tau)}{\sqrt n}\Big) -f(0)\right|\\
    &\le \sup_{g\in BL_1(\mathbb R)}\left|\E_Wg(U_n)-\E g(\mathbb G_Y(\tau))\right| +\E\frac{|\mathbb G_Y(\tau)|}{\sqrt n}\overset{\P^*_X}{\longrightarrow}0.
\end{align*}
Therefore, $V_n\overunderset{\P}{W}{\rightsquigarrow}0$. Now take $h(x)\coloneqq 1\land |x|$, which belongs to $BL_1(\mathbb R)$. For every $\eta>0$, $\mathbbm 1\{|x|>\eta\}\le (\eta\land 1)^{-1}h(x)$, whence
\begin{equation*}
    \P_W(|V_n|>\eta)\le \frac{\E_Wh(V_n)}{1\land \eta}\overset{\P_X}{\longrightarrow}\frac{h(0)}{1\land \eta}=0.
\end{equation*}
It follows that
\begin{align*}
    \P_W(\bar Y_n^W(\tau)<y_\tau/2)&\le  \P_W(|V_n|>y_\tau/4)+\mathbbm 1\{|\bar Y_n(\tau)-y_\tau|\ge y_\tau/4\}\overset{\P_X}{\longrightarrow}0.
\end{align*}
Combining the preceding displays yields
\begin{equation*}
    \P_W((A_n^W)^*>\varepsilon)\overset{\P_X}{\longrightarrow}0.
\end{equation*}

We next treat $B_n^W$. Set
\begin{equation*}
M_n\coloneqq \frac1n\sum_{i=1}^n \delta_i\mathbbm 1\{Z_i>\tau\} \qquad\text{and}\qquad M_n^W\coloneqq \frac1n\sum_{i=1}^n W_{ni}\delta_i\mathbbm 1\{Z_i>\tau\},
\end{equation*}
and let $p_\tau \coloneqq \E M_n=\P(\tau<T\le C)$. If $p_\tau=0$, then there is no uncensored event after $\tau$ almost surely, and hence $B_n^W=0$ almost surely. In that case, the box calibration is not active on the tail interval $[E_{n,m_n},\tau]$ and the geometric correction has no effect. Assume $p_\tau>0$. On the event $\{M_n^W>0\}$, the time $E_{n,m_n+1}$ is finite and every subject with $\tau<T_i\le C_i$ is at risk at time $E_{n,m_n+1}$. Therefore, $\bar Y_n^W(E_{n,m_n+1})\ge M_n^W$. It follows that
\begin{equation*}
    B_n^W \le \frac{W_{n,\pi(m_n+1)}}{\sqrt n\,\bar Y_n^W(E_{n,m_n+1})}\le \frac{\max_{1\le i\le n}W_{ni}/\sqrt n}{M_n^W}
\end{equation*}
where $M_n^W$ is $(\mathscr X^\infty\otimes \mathscr W)$-measurable. Hence
\begin{align*}
\P_W((B_n^W)^*>\varepsilon) &\le \P_W\!\left(M_n^W<\frac{M_n}{2}\right) + \P_W\!\left(\max_{1\le i\le n}\frac{W_{ni}}{\sqrt n}>\frac{\varepsilon M_n}{2}\right)\\
&\le \P_W\!\left(M_n^W<\frac{M_n}{2}\right) +\mathbbm 1\!\left\{M_n<\frac{p_\tau}{2}\right\}+
\P_W\!\left(\max_{1\le i\le n}\frac{W_{ni}}{\sqrt n}>\frac{\varepsilon p_\tau}{4}\right).
\end{align*}
Now $M_n\to p_\tau$ $\P_X$-almost surely by the strong law, the last term is $o(1)$ by
Condition~1, and
\begin{equation*}
    M_n^W-M_n=\frac1n\sum_{i=1}^n(W_{ni}-1)\delta_i\mathbbm 1\{Z_i>\tau\}\overunderset{\P}{W}{\rightsquigarrow}0
\end{equation*}
by the same argument used for $\bar Y_n^W(\tau)-\bar Y_n(\tau)$. Consequently,
\begin{equation*}
    \P_W((B_n^W)^*>\varepsilon)\overset{\P_X}{ \longrightarrow}0.
\end{equation*}
\ 
\end{proof}

\section{More simulation results}
\begin{table}[htbp]
\centering
\caption{
    Simulation summary of empirical coverages for the six bootstrap-based methods and the Hall--Wellner band under light (L, $\sim$20\%) and heavy (H, $\sim$50\%) censoring over $[0,1]$, at the nominal 99\% level.
    The 95\% Monte Carlo reference interval, reported in percent, is $[98.938,99.062]$.
}
\label{tab:alpha001}
\small
\setlength{\tabcolsep}{4pt}
\begin{tabular}{l *{12}{c}}
% \toprule
& \multicolumn{2}{c}{$n=15$} & \multicolumn{2}{c}{$n=25$} & \multicolumn{2}{c}{$n=50$} & \multicolumn{2}{c}{$n=100$} & \multicolumn{2}{c}{$n=200$} & \multicolumn{2}{c}{$n=500$} \\
% \cmidrule(lr){2-3}\cmidrule(lr){4-5}\cmidrule(lr){6-7}\cmidrule(lr){8-9}\cmidrule(lr){10-11}\cmidrule(lr){12-13}
Method & L & H & L & H & L & H & L & H & L & H & L & H \\
% \midrule

\multicolumn{13}{@{}l}{\textit{Exponential}(1): $H_0(t)=t$} \\
Wild--Grid & $97.60$ & $95.79$ & $98.02$ & $97.06$ & $98.41$ & $97.96$ & $98.65$ & $98.39$ & $98.70$ & $98.52$ & $98.77$ & $98.70$ \\
Wild--Box & $99.30$ & $99.49$ & $99.25$ & $99.63$ & $99.26$ & $99.44$ & $99.18$ & $99.33$ & $99.12$ & $99.24$ & $98.99$ & $99.12$ \\
Weird--Grid & $94.50$ & $86.39$ & $96.34$ & $92.99$ & $97.59$ & $96.41$ & $98.26$ & $97.62$ & $98.52$ & $98.10$ & $98.70$ & $98.53$ \\
Weird--Box & $98.29$ & $98.38$ & $98.48$ & $98.90$ & $98.74$ & $98.87$ & $98.88$ & $98.92$ & $98.92$ & $98.94$ & $98.92$ & $98.93$ \\
XBoot--Grid & $94.71$ & $86.19$ & $96.67$ & $93.25$ & $97.81$ & $96.74$ & $98.39$ & $97.91$ & $98.57$ & $98.27$ & $98.72$ & $98.60$ \\
XBoot--Box & $98.42$ & $98.34$ & $98.73$ & $98.96$ & $99.00$ & $99.07$ & $99.03$ & $99.14$ & $99.04$ & $99.11$ & $98.95$ & $99.02$ \\
Hall--Wellner & $94.68$ & $90.84$ & $96.11$ & $93.16$ & $97.36$ & $95.56$ & $98.14$ & $97.05$ & $98.58$ & $97.96$ & $98.80$ & $98.55$ \\
% HW & $94.64$ & $90.70$ & $96.10$ & $93.17$ & $97.26$ & $95.56$ & $98.08$ & $97.11$ & $98.52$ & $97.94$ & $98.81$ & $98.52$ \\
% logHW & $97.98$ & $97.97$ & $98.37$ & $98.37$ & $98.84$ & $98.83$ & $98.99$ & $98.93$ & $99.00$ & $99.03$ & $99.02$ & $99.00$ \\
% \midrule

\multicolumn{13}{@{}l}{\textit{Weibull}$(0.5,0.5)$: $H_0(t)=\sqrt{2t}$} \\
Wild--Grid & $97.45$ & $93.51$ & $97.96$ & $95.49$ & $98.37$ & $97.17$ & $98.60$ & $97.98$ & $98.73$ & $98.42$ & $98.78$ & $98.63$ \\
Wild--Box & $99.26$ & $98.09$ & $99.21$ & $99.16$ & $99.27$ & $99.51$ & $99.23$ & $99.47$ & $99.15$ & $99.33$ & $99.06$ & $99.22$ \\
Weird--Grid & $90.12$ & $76.73$ & $95.43$ & $84.05$ & $97.45$ & $92.60$ & $98.15$ & $96.57$ & $98.49$ & $97.81$ & $98.69$ & $98.31$ \\
Weird--Box & $98.07$ & $95.52$ & $98.40$ & $97.70$ & $98.73$ & $98.74$ & $98.89$ & $98.94$ & $98.97$ & $98.95$ & $98.97$ & $98.99$ \\
XBoot--Grid & $89.55$ & $72.41$ & $95.30$ & $82.25$ & $97.44$ & $92.44$ & $98.22$ & $96.61$ & $98.52$ & $97.94$ & $98.69$ & $98.39$ \\
XBoot--Box & $98.06$ & $95.21$ & $98.47$ & $97.59$ & $98.91$ & $98.73$ & $99.03$ & $99.10$ & $99.06$ & $99.13$ & $99.00$ & $99.08$ \\
Hall--Wellner & $93.36$ & $86.06$ & $95.11$ & $89.70$ & $96.79$ & $93.53$ & $97.79$ & $95.89$ & $98.33$ & $97.39$ & $98.69$ & $98.32$ \\
% HW & $93.29$ & $85.69$ & $95.08$ & $89.40$ & $96.82$ & $93.44$ & $97.78$ & $95.88$ & $98.38$ & $97.43$ & $98.74$ & $98.40$ \\
% logHW & $97.95$ & $97.78$ & $98.49$ & $98.20$ & $98.79$ & $98.58$ & $98.95$ & $98.74$ & $99.00$ & $98.93$ & $99.01$ & $98.95$ \\
% \midrule

\multicolumn{13}{@{}l}{\textit{Weibull}$(2,1/\Gamma(3/2))$: $H_0(t)=\Gamma(3/2)^2 t^2$} \\
Wild--Grid & $97.55$ & $96.33$ & $98.10$ & $97.37$ & $98.42$ & $98.04$ & $98.55$ & $98.44$ & $98.67$ & $98.55$ & $98.79$ & $98.73$ \\
Wild--Box & $99.28$ & $99.64$ & $99.26$ & $99.59$ & $99.18$ & $99.38$ & $99.14$ & $99.29$ & $99.08$ & $99.16$ & $99.02$ & $99.08$ \\
Weird--Grid & $94.78$ & $91.20$ & $96.37$ & $94.75$ & $97.58$ & $96.82$ & $98.17$ & $97.77$ & $98.52$ & $98.25$ & $98.71$ & $98.61$ \\
Weird--Box & $98.28$ & $98.90$ & $98.50$ & $98.92$ & $98.71$ & $98.81$ & $98.80$ & $98.88$ & $98.89$ & $98.89$ & $98.95$ & $98.95$ \\
XBoot--Grid & $95.16$ & $91.68$ & $96.97$ & $95.38$ & $97.94$ & $97.29$ & $98.34$ & $98.07$ & $98.57$ & $98.37$ & $98.76$ & $98.68$ \\
XBoot--Box & $98.56$ & $98.97$ & $98.83$ & $99.11$ & $98.97$ & $99.09$ & $99.01$ & $99.12$ & $98.98$ & $99.05$ & $98.98$ & $99.05$ \\
Hall--Wellner & $94.86$ & $92.57$ & $96.45$ & $94.44$ & $97.54$ & $96.40$ & $98.22$ & $97.61$ & $98.62$ & $98.22$ & $98.82$ & $98.72$ \\
% HW & $95.04$ & $92.63$ & $96.32$ & $94.69$ & $97.55$ & $96.44$ & $98.22$ & $97.62$ & $98.63$ & $98.21$ & $98.85$ & $98.74$ \\
% logHW & $97.81$ & $97.82$ & $98.35$ & $98.33$ & $98.79$ & $98.84$ & $98.94$ & $99.03$ & $99.02$ & $99.06$ & $99.02$ & $98.98$ \\
% \midrule

\multicolumn{13}{@{}l}{\textit{Piecewise-constant hazard}: $h_0(t)=0.5\,\mathbbm{1}\{t\le \tau_0\}+2\,\mathbbm{1}\{t>\tau_0\},\,\tau_0=2\log(3/2)$} \\
Wild--Grid & $97.58$ & $96.34$ & $98.06$ & $97.30$ & $98.37$ & $98.09$ & $98.56$ & $98.46$ & $98.71$ & $98.64$ & $98.79$ & $98.79$ \\
Wild--Box & $99.28$ & $99.64$ & $99.27$ & $99.58$ & $99.20$ & $99.39$ & $99.12$ & $99.29$ & $99.10$ & $99.21$ & $98.99$ & $99.11$ \\
Weird--Grid & $94.79$ & $91.34$ & $96.41$ & $94.79$ & $97.55$ & $96.83$ & $98.22$ & $97.86$ & $98.54$ & $98.34$ & $98.74$ & $98.66$ \\
Weird--Box & $98.27$ & $98.89$ & $98.45$ & $98.87$ & $98.68$ & $98.84$ & $98.81$ & $98.92$ & $98.93$ & $98.94$ & $98.92$ & $98.98$ \\
XBoot--Grid & $95.22$ & $91.74$ & $97.00$ & $95.37$ & $97.90$ & $97.34$ & $98.36$ & $98.12$ & $98.62$ & $98.47$ & $98.75$ & $98.74$ \\
XBoot--Box & $98.54$ & $98.98$ & $98.82$ & $99.03$ & $98.97$ & $99.11$ & $99.01$ & $99.16$ & $99.05$ & $99.13$ & $98.98$ & $99.07$ \\
Hall--Wellner & $95.07$ & $92.60$ & $96.41$ & $94.65$ & $97.57$ & $96.48$ & $98.29$ & $97.65$ & $98.69$ & $98.33$ & $98.85$ & $98.75$ \\
% HW & $95.04$ & $92.53$ & $96.47$ & $94.62$ & $97.59$ & $96.46$ & $98.24$ & $97.60$ & $98.52$ & $98.19$ & $98.87$ & $98.73$ \\
% logHW & $97.78$ & $97.88$ & $98.35$ & $98.45$ & $98.76$ & $98.83$ & $98.91$ & $98.95$ & $98.90$ & $98.98$ & $99.01$ & $99.09$ \\
% \bottomrule
\end{tabular}
\end{table}
\begin{table}[htbp]
\centering
\caption{
    Simulation summary of empirical coverages for the six bootstrap-based methods and the Hall--Wellner band under light (L, $\sim$20\%) and heavy (H, $\sim$50\%) censoring over $[0,1]$, at the nominal 90\% level.
    The 95\% Monte Carlo reference interval, reported in percent, is $[89.814, 90.186]$.
}
\label{tab:alpha010}
\small
\setlength{\tabcolsep}{4pt}
\begin{tabular}{l *{12}{c}}
% \toprule
& \multicolumn{2}{c}{$n=15$} & \multicolumn{2}{c}{$n=25$} & \multicolumn{2}{c}{$n=50$} & \multicolumn{2}{c}{$n=100$} & \multicolumn{2}{c}{$n=200$} & \multicolumn{2}{c}{$n=500$} \\
% \cmidrule(lr){2-3}\cmidrule(lr){4-5}\cmidrule(lr){6-7}\cmidrule(lr){8-9}\cmidrule(lr){10-11}\cmidrule(lr){12-13}
Method & L & H & L & H & L & H & L & H & L & H & L & H \\
% \midrule

\multicolumn{13}{@{}l}{\textit{Exponential}(1): $H_0(t)=t$} \\
Wild--Grid & $81.64$ & $77.48$ & $83.64$ & $80.63$ & $85.79$ & $83.61$ & $87.00$ & $85.62$ & $88.06$ & $86.69$ & $88.72$ & $87.99$ \\
Wild--Box & $92.23$ & $94.13$ & $91.85$ & $93.93$ & $90.96$ & $92.80$ & $90.46$ & $91.60$ & $90.28$ & $90.61$ & $89.99$ & $90.29$ \\
Weird--Grid & $78.27$ & $69.16$ & $81.78$ & $76.53$ & $84.99$ & $81.99$ & $86.68$ & $84.93$ & $87.87$ & $86.46$ & $88.63$ & $87.89$ \\
Weird--Box & $90.09$ & $90.86$ & $90.59$ & $91.91$ & $90.35$ & $91.78$ & $90.19$ & $91.09$ & $90.17$ & $90.36$ & $89.92$ & $90.17$ \\
XBoot--Grid & $73.51$ & $62.65$ & $78.83$ & $71.50$ & $83.50$ & $79.33$ & $85.87$ & $83.55$ & $87.49$ & $85.73$ & $88.52$ & $87.62$ \\
XBoot--Box & $88.37$ & $88.91$ & $89.01$ & $90.48$ & $89.36$ & $90.59$ & $89.59$ & $90.27$ & $89.79$ & $89.83$ & $89.83$ & $89.96$ \\
Hall--Wellner & $85.14$ & $79.44$ & $87.07$ & $82.65$ & $88.93$ & $85.96$ & $89.69$ & $87.83$ & $90.05$ & $89.14$ & $90.14$ & $89.75$ \\
% HW & $85.12$ & $79.34$ & $87.01$ & $82.47$ & $88.91$ & $85.69$ & $89.88$ & $88.00$ & $90.25$ & $89.07$ & $90.28$ & $89.88$ \\
% logHW & $90.22$ & $89.51$ & $90.40$ & $90.58$ & $90.70$ & $90.99$ & $90.76$ & $90.90$ & $90.46$ & $90.71$ & $90.40$ & $90.56$ \\
% \midrule

\multicolumn{13}{@{}l}{\textit{Weibull}$(0.5,0.5)$: $H_0(t)=\sqrt{2t}$} \\
Wild--Grid & $81.58$ & $73.77$ & $83.67$ & $77.86$ & $85.67$ & $81.73$ & $86.97$ & $83.83$ & $87.70$ & $85.69$ & $88.65$ & $87.25$ \\
Wild--Box & $92.13$ & $90.24$ & $92.35$ & $92.69$ & $91.87$ & $93.56$ & $91.05$ & $92.78$ & $90.37$ & $91.76$ & $90.19$ & $90.66$ \\
Weird--Grid & $73.87$ & $51.04$ & $79.90$ & $63.97$ & $84.17$ & $76.38$ & $86.34$ & $81.82$ & $87.41$ & $84.86$ & $88.56$ & $86.92$ \\
Weird--Box & $88.42$ & $85.43$ & $90.32$ & $88.87$ & $90.84$ & $91.20$ & $90.58$ & $91.59$ & $90.19$ & $91.16$ & $90.10$ & $90.44$ \\
XBoot--Grid & $68.13$ & $43.68$ & $75.75$ & $58.05$ & $82.07$ & $71.65$ & $85.36$ & $79.21$ & $86.92$ & $83.53$ & $88.34$ & $86.40$ \\
XBoot--Box & $86.14$ & $82.77$ & $88.39$ & $86.58$ & $89.43$ & $89.45$ & $89.77$ & $90.27$ & $89.69$ & $90.28$ & $89.89$ & $90.01$ \\
Hall--Wellner & $82.45$ & $73.06$ & $85.19$ & $77.30$ & $87.47$ & $82.04$ & $88.96$ & $85.38$ & $89.62$ & $87.52$ & $90.07$ & $89.07$ \\
% HW & $82.37$ & $72.60$ & $85.03$ & $76.89$ & $87.54$ & $81.99$ & $89.01$ & $85.37$ & $89.48$ & $87.48$ & $89.92$ & $89.02$ \\
% logHW & $90.05$ & $88.75$ & $90.42$ & $89.08$ & $90.71$ & $89.62$ & $90.72$ & $89.84$ & $90.50$ & $90.14$ & $90.31$ & $90.24$ \\
% \midrule

\multicolumn{13}{@{}l}{\textit{Weibull}$(2,1/\Gamma(3/2))$: $H_0(t)=\Gamma(3/2)^2 t^2$} \\
Wild--Grid & $81.38$ & $78.50$ & $83.65$ & $81.21$ & $85.74$ & $84.20$ & $87.23$ & $85.98$ & $88.06$ & $87.14$ & $88.74$ & $88.27$ \\
Wild--Box & $92.08$ & $94.63$ & $91.29$ & $93.43$ & $90.65$ & $91.93$ & $90.36$ & $90.97$ & $90.07$ & $90.30$ & $89.98$ & $90.17$ \\
Weird--Grid & $79.35$ & $74.63$ & $82.49$ & $79.29$ & $85.25$ & $83.36$ & $86.98$ & $85.63$ & $87.98$ & $87.00$ & $88.68$ & $88.22$ \\
Weird--Box & $90.32$ & $92.47$ & $90.34$ & $92.07$ & $90.20$ & $91.23$ & $90.14$ & $90.74$ & $90.01$ & $90.17$ & $89.92$ & $90.09$ \\
XBoot--Grid & $75.49$ & $68.56$ & $80.05$ & $75.40$ & $83.98$ & $81.36$ & $86.33$ & $84.63$ & $87.67$ & $86.49$ & $88.56$ & $87.98$ \\
XBoot--Box & $88.92$ & $91.15$ & $89.09$ & $90.86$ & $89.38$ & $90.26$ & $89.63$ & $90.02$ & $89.72$ & $89.79$ & $89.79$ & $89.92$ \\
Hall--Wellner & $86.05$ & $82.62$ & $87.87$ & $85.13$ & $89.52$ & $87.66$ & $90.06$ & $89.10$ & $90.27$ & $89.73$ & $90.15$ & $90.11$ \\
% HW & $86.28$ & $82.49$ & $87.95$ & $85.15$ & $89.52$ & $87.61$ & $90.01$ & $89.12$ & $90.42$ & $89.72$ & $90.29$ & $90.08$ \\
% logHW & $89.80$ & $89.83$ & $90.16$ & $90.70$ & $90.59$ & $91.16$ & $90.56$ & $91.07$ & $90.42$ & $90.71$ & $90.31$ & $90.58$ \\
% \midrule

\multicolumn{13}{@{}l}{\textit{Piecewise-constant hazard}: $h_0(t)=0.5\,\mathbbm{1}\{t\le \tau_0\}+2\,\mathbbm{1}\{t>\tau_0\},\,\tau_0=2\log(3/2)$} \\
Wild--Grid & $81.55$ & $78.47$ & $83.67$ & $81.53$ & $85.87$ & $84.02$ & $87.24$ & $86.02$ & $88.25$ & $87.35$ & $88.80$ & $88.56$ \\
Wild--Box & $92.15$ & $94.57$ & $91.25$ & $93.42$ & $90.67$ & $91.83$ & $90.40$ & $91.04$ & $90.33$ & $90.63$ & $90.02$ & $90.44$ \\
Weird--Grid & $79.53$ & $74.73$ & $82.57$ & $79.51$ & $85.34$ & $83.21$ & $87.06$ & $85.66$ & $88.12$ & $87.25$ & $88.73$ & $88.55$ \\
Weird--Box & $90.37$ & $92.35$ & $90.30$ & $92.05$ & $90.26$ & $91.16$ & $90.23$ & $90.75$ & $90.25$ & $90.46$ & $89.98$ & $90.37$ \\
XBoot--Grid & $75.54$ & $68.73$ & $80.06$ & $75.76$ & $84.06$ & $81.19$ & $86.41$ & $84.66$ & $87.80$ & $86.73$ & $88.59$ & $88.31$ \\
XBoot--Box & $88.97$ & $91.03$ & $89.03$ & $90.87$ & $89.43$ & $90.13$ & $89.72$ & $90.03$ & $89.97$ & $90.10$ & $89.85$ & $90.17$ \\
Hall--Wellner & $86.18$ & $82.66$ & $88.08$ & $85.23$ & $89.54$ & $87.61$ & $90.28$ & $89.22$ & $90.39$ & $89.74$ & $90.24$ & $90.19$ \\
% HW & $86.18$ & $82.65$ & $88.06$ & $85.18$ & $89.59$ & $87.65$ & $90.08$ & $88.97$ & $90.23$ & $89.65$ & $90.17$ & $90.20$ \\
% logHW & $89.43$ & $89.68$ & $90.15$ & $90.74$ & $90.56$ & $90.97$ & $90.38$ & $90.92$ & $90.29$ & $90.74$ & $90.33$ & $90.64$ \\
% \bottomrule
\end{tabular}
\end{table}
%\vspace*{-10pt}

\section{A two-channel decomposition of the box critical-value bias}
\label{ssec:decomp}

Section~5.3 of the main text reports the scaled signed coverage deviation of the three box-calibrated bands (Figure~2). Here we decompose that finite-sample behaviour into a distributional component and a box-inflation component, and show that the geometric content the box statistic adds beyond grid calibration is, at the coverage-setting quantile, near zero. We use the same data-generating mechanisms, censoring levels, and Monte Carlo budget as in Section~5 of the main text ($R=100{,}000$ replications, $B=2{,}000$ bootstrap resamples, upper bound $\tau=1$), under heavy censoring and over the sample-size grid of Figure~2 of the main text. Throughout this section critical values are reported on the $\sqrt n$ scale, on which they are $O(1)$ and comparable across $n$.

Because $H_0$ is known in simulation, the oracle critical values for the two calibration geometries are computable. Let $\widehat\Delta_{n,\mathrm{grid}}^W$ and $\widehat\Delta_{n,\mathrm{box}}^W$ be the grid and box discrepancies of equation (13) and equation (15), and let $\hat q^{W}_{n,\mathrm{grid}}(1-\alpha)$ and $\hat q^{W}_{n,\mathrm{box}}(1-\alpha)$ be their conditional $(1-\alpha)$ quantiles, that is, the half-widths of the bands equation (12) and equation (17) of the main text. Define the oracle grid and box critical values
\begin{equation*}
c_{\mathrm{grid}}(\alpha)=Q_{1-\alpha}\!\Bigl(\sqrt n\max_{1\le i\le m_n}\bigl|\widehat H_n(E_{ni})-H_0(E_{ni})\bigr|\Bigr),\qquad
c_{\mathrm{box}}(\alpha)=Q_{1-\alpha}\!\bigl(\sqrt n\,\Delta_n\bigr),
\end{equation*}
where $E_{ni}$ are the uncensored event times of equation (5), $\Delta_n=\|\widehat H_n-H_0\|_{\mathcal T}$ is the supremum error of equation (14), and $Q_{1-\alpha}$ denotes the population $(1-\alpha)$ quantile, here estimated by an independent Monte Carlo of $10^{6}$ datasets. Writing $\bar q^{s}_{\mathrm{grid}}=\E_P[\sqrt n\,\hat q^{W}_{n,\mathrm{grid}}(1-\alpha)]$ and $\bar q^{s}_{\mathrm{box}}$ analogously for a scheme $s\in\{\mathrm{XBoot},\mathrm{Wild},\mathrm{Weird}\}$, the box critical-value bias admits the exact decomposition
\begin{equation}
\underbrace{\bar q^{s}_{\mathrm{box}}-c_{\mathrm{box}}}_{b^{s}\ \text{(total)}}
=\underbrace{\bigl(\bar q^{s}_{\mathrm{grid}}-c_{\mathrm{grid}}\bigr)}_{D^{s}\ \text{(distributional)}}
+\underbrace{\bigl\{(\bar q^{s}_{\mathrm{box}}-\bar q^{s}_{\mathrm{grid}})-(c_{\mathrm{box}}-c_{\mathrm{grid}})\bigr\}}_{G^{s}\ \text{(box inflation)}},
\label{eq:decomp}
\end{equation}
in which $D^{s}$ records how far grid calibration alone falls short of its oracle and $G^{s}$ records the box correction net of the geometric correction already present in the oracle. 
By the first-order equivalence of grid and box calibration (Theorem~2 of the main text), $c_{\mathrm{box}}-c_{\mathrm{grid}}\to 0$; in the present experiment its median is $0.001$, $0.029$, and $0.052$ at $\alpha=0.01,0.05,0.10$, so $G^{s}$ is predominantly the bootstrap box correction itself instead of a matching of a large oracle gap.

\begin{figure}[ht!]
    \centering
    \includegraphics[width=\linewidth]{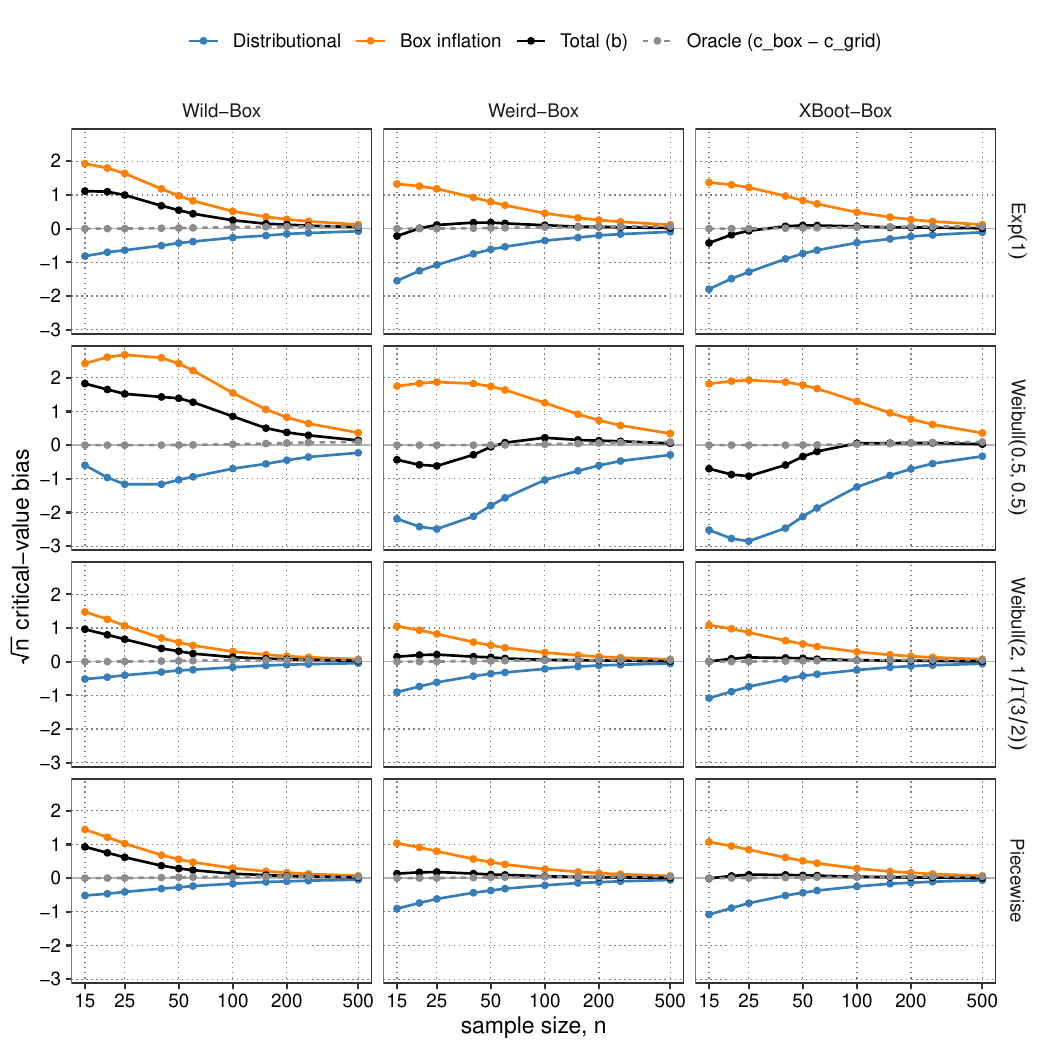}
    \caption{
        Decomposition \eqref{eq:decomp} of the box critical-value bias on the $\sqrt n$ scale, under heavy censoring at $\alpha=0.05$.
        Columns are the three box-calibrated schemes, ordered as in Figure~2 of the main text; rows are the four hazard shapes.
        Within each panel the lines show the distributional channel $D$, the box inflation $G$, their sum the total bias $b=D+G$, and the oracle box correction $c_{\mathrm{box}}-c_{\mathrm{grid}}$ (dashed, near zero), which is the true geometric content the box statistic adds.
        For XBoot--Box the total lies near zero across $n$; for Wild--Box it remains positive.
    }
    \label{fig:decomp}
\end{figure}

Figure~\ref{fig:decomp} displays $D^{s}$, $G^{s}$, and their sum $b^{s}$ against $n$ for the three schemes at $\alpha=0.05$. Three features are visible. First, $D^{s}<0$ for every scheme: grid calibration undercovers, and most so for the exchangeable bootstrap, whose beta jump law equation (7) has the smallest conditional variance and hence the narrowest grid band. Second, $G^{s}>0$ for every scheme and is largest for the wild bootstrap. Third, the two channels combine differently across schemes: for XBoot--Box the deficit and the correction are of comparable size and largely cancel (averaged over the four hazards under heavy censoring, $\bar D\approx-0.82$, $\bar G\approx+0.76$, $\bar b\approx-0.06$), whereas for Wild--Box the correction exceeds the smaller deficit ($\bar D\approx-0.40$, $\bar G\approx+0.95$, $\bar b\approx+0.55$). 
We summarise the per-scheme offsets by the mean of per-cell ratios, $\kappa^{s}=G^{s}/(-D^{s})$, we obtain $\kappa\approx1.07$, $1.19$, and $2.14$ for XBoot, Weird, and Wild, respectively.

Equation \eqref{eq:decomp} accounts for the ranking reversal of Section~5.2 of the main text. The box correction is a finite-sample device whose magnitude is set by the conditional variance of the bootstrap jump sizes; for the exchangeable bootstrap that magnitude is close to its grid-calibration deficit, so the two channels offset.

\section{Grid thinning and the dimension of the maximum}
\label{ssec:thinning}

The distributional channel $D^{s}<0$ in \eqref{eq:decomp} states that the bootstrap's maximum over the event grid is, on average, smaller than the oracle. The likely explanation is a high-dimensional-maximum effect: the grid critical value is the $(1-\alpha)$ quantile of a maximum over many points, the event grid of equation (13), and the bootstrap approximation of the tail of such a maximum degrades with the number of points, by analogy with the logarithmic dimension factor in high-dimensional central-limit theory \citep{chernozhukov2017central,koike2026high}. In the natural experiment the number of event points and the sample size increase together, so the two are confounded. We therefore hold $n$ fixed and vary only the number of points entering the maximum.

We use the data-generating mechanisms of Section~5 of the main text under heavy censoring, with $R=100{,}000$ datasets and $B=2{,}000$ resamples, for three cells: $\mathrm{Exponential}(1)$ at $n=100$ (primary) and, for robustness, $\mathrm{Exponential}(1)$ at $n=50$ and the decreasing-hazard $\mathrm{Weibull}(0.5,0.5)$ at $n=100$. For a fixed dataset we form the Nelson--Aalen estimator and its event grid equation (5) as usual, and then, before calibrating, retain only a subset of the grid points. Two thinning rules are used: an endpoint-preserving rule that keeps roughly $1/k$ of the points spread evenly in rank while always retaining the first and last point, so that the high-variance right endpoint near $\tau$ is kept; and a naive rule that keeps every $k$-th point from the left and usually drops that endpoint. On each thinned grid we recompute, on the same datasets, both $c_{\mathrm{grid}}$ and $\bar q^{s}_{\mathrm{grid}}$, and form $D^{s}$ as in \eqref{eq:decomp}; we plot $D^{s}$ against $\bar d$, the mean number of retained points. At $k=1$ the two rules coincide, and the resulting $D^{s}$ reproduces the full-grid values of Section~\ref{ssec:decomp} above to Monte Carlo error.

\begin{figure}[ht!]
    \centering
    \includegraphics[width=\linewidth]{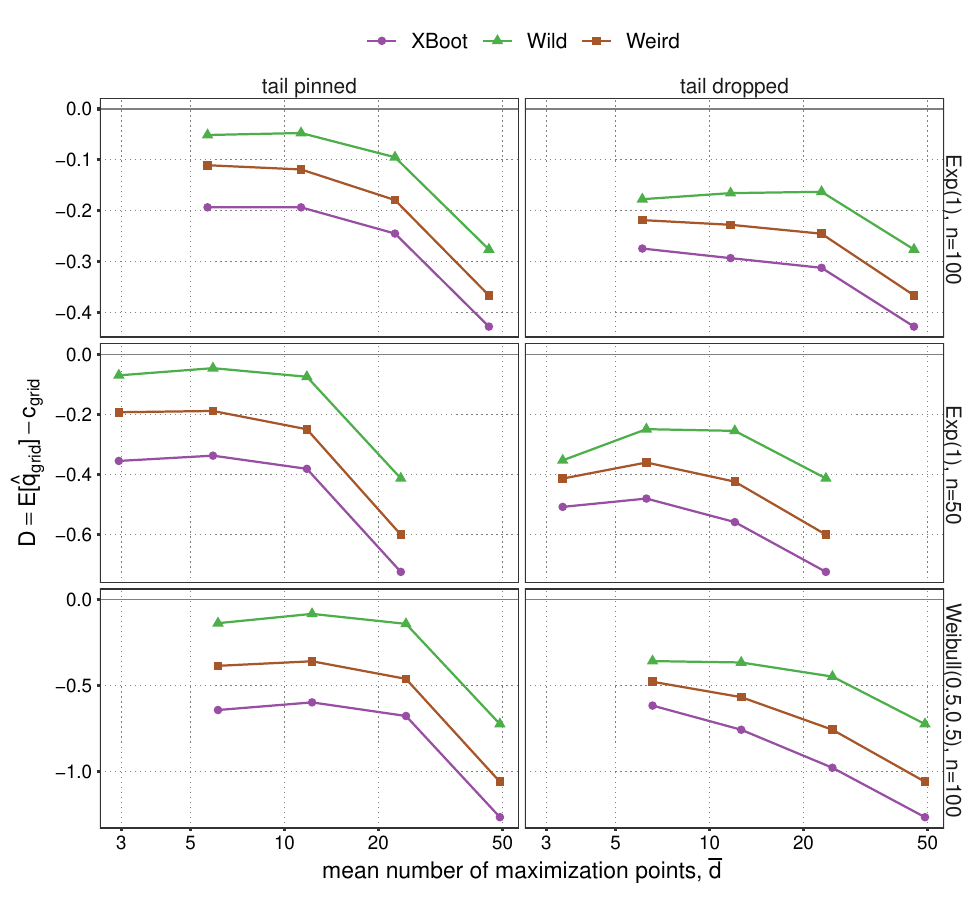}
    \caption{
        Grid thinning at fixed sample size.
        The distributional channel $D=\bar q^{s}_{\mathrm{grid}}-c_{\mathrm{grid}}$, on the $\sqrt n$ scale at $\alpha=0.05$ under heavy censoring, against the mean number $\bar d$ of retained maximization points, for the three schemes.
        Rows are the three cells; the left column keeps the endpoints (tail pinned) and the right column drops the tail.
        At the full grid the two columns coincide. As $\bar d$ falls, $|D|$ decreases for every scheme.
    }
    \label{fig:thinning}
\end{figure}

Figure~\ref{fig:thinning} shows the result at $\alpha=0.05$. 
As the grid is thinned, $|D^{s}|$ decreases toward zero for all three schemes, and grows as more points are retained. 
Two further features support reading this as a dimensional effect rather than an artefact of which points are kept. The three schemes attenuate in near-parallel, with similar slopes but differing levels with each cell (for $\mathrm{Exponential}(1)$, $n=100$, the slopes against $\bar d$ are $-0.109$, $-0.104$, and $-0.120$ for XBoot, Wild, and Weird).
This is consistent with a scheme-specific per-increment under-dispersion, governed by the conditional-variance ordering of Section~3 of the main text, amplified by a common factor that depends on the size of the maximization grid. 
In addition, the endpoint-preserving and naive rules attenuate in the same direction, so it is the number of points, and not the retention of the high-variance right endpoint, that moves $D^{s}$.

The grid-calibration deficit thus grows with the number of maximization points, as the high-dimensional-maximum account predicts. The scheme-dependent levels show that the resampling scheme also matters, so the grid size is not the only driver of $D^{s}$.

\bibliographystyle{biometrika}
\bibliography{boxcalibration.bib}